\documentclass[
 reprint,
 amsmath,amssymb,
 aps,
]{revtex4-2}
\newcommand{\Aphys}{A_{\mathrm{phys}}}
\newcommand{\Axy}{A_{xy}}
\newcommand{\QL}{Q_{\mathrm L}}
\newcommand{\Qext}{Q_{\mathrm{ext}}}
\newcommand{\Erec}{E_{\mathrm{rec},y}}
\newcommand{\Pin}{P_{\mathrm{in}}}
\newcommand{\Pout}{P_{\mathrm{out}}}
\newcommand{\Pmode}{P_{\mathrm{mode}}}
\newcommand{\Ploss}{P_{\mathrm{loss}}}
\newcommand{\gagg}{g_{a\gamma\gamma}}
\newcommand{\CB}{\mathcal{C}_{B}}
\newcommand{\Zport}{Z_{\mathrm{port}}}
\usepackage{graphicx}
\usepackage{bm}
\usepackage{booktabs}
\usepackage[colorlinks=true,linkcolor=blue,citecolor=blue]{hyperref}
\begin{document}
\preprint{APS/123-QED}
\title{The DALI Haloscope:\\ A Magnetized Phased Array Coupled to a Semi-Open Fabry--Pérot Resonator}

\author{Javier De Miguel$^{1,2,3}$}
 \email{jdemiguel@iac.es}


\author{Antonios Gardikiotis$^{4,5}$}%
\author{Eduardo David Gonz\'alez-Carretero$^{1}$}%
\author{Elvio Hern\'andez-Su\'arez$^{1}$}%
\author{Roger J. Hoyland$^{1}$}%
\author{Enrique Joven$^{1}$}%
\author{Abaz Kryemadhi$^{6}$}%
\author{Marios Maroudas$^{7}$}%
\author{Chiko Otani$^{3}$}%
\author{J. Alberto Rubi\~no-Mart\'in$^{1,2}$}%
\author{Yannis K. Semertzidis$^{8}$}%
\author{Michael E. Tobar$^{9}$}%
\author{Konstantin Zioutas$^{4}$}%

\collaboration{The DALI Collaboration}

\affiliation{$^{1}$Instituto de Astrof\'isica de Canarias, E-38200 La Laguna, Tenerife, Spain}

\affiliation{$^{2}$Departamento de Astrof\'isica, Universidad de La Laguna, E-38206 La Laguna, Tenerife, Spain}

\affiliation{$^{3}$RIKEN Center for Advanced Photonics,
519-1399 Aramaki-Aoba, Aoba-ku, Sendai, Miyagi 980-0845, Japan}

\affiliation{$^{4}$Physics Department, University of Patras, GR 26504, Patras-Rio, Greece}

\affiliation{$^{5}$Institute of Quantum Computing and Quantum Technology, National Centre for Scientific Research ``Demokritos'', 153 41 Athens, Greece}

\affiliation{$^{6}$Computing, Math \& Physics, Messiah University, Mechanicsburg, PA 17055, USA}

\affiliation{$^{7}$Institute for Experimental Physics, University of Hamburg, 22761 Hamburg, Germany}

\affiliation{$^{8}$Innovative Solutions R\&D LLC, Stony Brook, NY 11790, USA}

\affiliation{$^{9}$Quantum Technologies and Dark Matter Research Lab, Department of Physics, University of Western Australia, Crawley, WA 6009, Australia.}

\date{\today}

\begin{abstract}
DALI is an axion haloscope consisting of a magnetized phased array backed by a conducting mirror and coupled in the near field to a semi-open Fabry--Pérot resonator. The applied static magnetic field makes both the dielectric interfaces and the conducting mirror sensitive to axion-induced electromagnetic conversion. Each interface acts as a radiating surface, and the resulting fields are projected onto the mode collected by the resonator and delivered to the receiver. We derive the detected power in terms of the power available from the magnetized mirror, the coherent contribution of the dielectric interfaces, the transverse mode overlap, the loaded quality factor, and the receiver-coupling coefficient.

This formulation separates two distinct enhancement mechanisms. The dielectric-interface emissions may add coherently, producing a dielectric boost, while the semi-open Fabry--Pérot resonator provides enhancement through resonant storage of the coupled electromagnetic field. These contributions depend differently on the dielectric thicknesses, spacings, and resonant mode, and should therefore be evaluated independently. The same assembly can operate between the limits of a coherently boosted dielectric haloscope and a mirror-sourced, resonantly enhanced Fabry--Pérot haloscope. This combined architecture offers a flexible approach to resonant axion searches at frequencies for which conventional closed microwave cavities become increasingly limited in conversion volume.

\end{abstract}

\maketitle

\section{Introduction}\label{sec:intro}

The axion was originally introduced as a solution to the
charge-conjugation and parity problem in quantum chromodynamics
(QCD)~\cite{PhysRevLett.38.1440}.  QCD
axions~\cite{PhysRevLett.40.223,PhysRevLett.40.279}, together with
axionlike particles not tied to the QCD solution---collectively denoted
as axions in this work---are well-motivated candidates for dark matter
(DM)~\cite{1933AcHPh...6..110Z,ABBOTT1983133,DINE1983137,PRESKILL1983127}.
The DM hypothesis is supported by a broad range of indirect astrophysical
observations~\cite{1970ApJ...159..379R}.  Their feeble interactions with
Standard Model particles motivate a diverse experimental programme aimed
at detecting axion DM; recent reviews can be found in
Refs.~\cite{Arza:2026rsl,Andrieu:2025xpv}.

The interaction relevant to haloscope searches is the axion--photon
coupling, described by the effective Lagrangian density
\begin{equation}
\mathcal{L} \supset g_{a\gamma\gamma} a \, \mathbf{E} \cdot \mathbf{B} \,,
\end{equation}
where $g_{a\gamma\gamma}$ is the axion--photon coupling constant, $a$
denotes the axion field, \(\mathbf{E}\) is the electromagnetic field and \(\mathbf{B}\) is the externally applied static magnetic field, which providesprovides the virtual photon required for Primakoff conversion~\cite{Primakoff:1951iae}.  Haloscopes,
first proposed by Sikivie~\cite{1983PhRvL..51.1415S}, exploit this
interaction by converting virialized Galactic axions into microwave
photons within a magnetized resonator, resonant enhancement compensating
for the exceptionally small expected signal power.

Conventional cavity haloscopes become increasingly difficult to scale to
high frequencies: maintaining electromagnetic coherence requires
progressively smaller resonator dimensions, reducing the available
detection volume and therefore the achievable sensitivity.  The
Dark-photons \& Axion-Like particles Interferometer
(DALI)~\cite{DeMiguel2021,DeMiguel:2023nmz,Cabrera2023qkt,2024JInst19P1022H,PhysRevD.110.072013,DeMiguel:2024cwb,DeMiguel:2026mvi}
takes a different route.  It is a magnetized phased array coupled, in the radiative
near field, to a semi-open multilayer Fabry--P\'erot resonator, and provides a
complementary strategy for high-frequency resonant axion searches.  Its
implementation relies on mature and readily available components,
including solenoid-type superconducting magnets widely used in the
medical sector, which reduces technical risk and facilitates
cost-effective development. We derive the power delivered at the receiver, factor by factor, and
identify which properties of the eigenmode enter each factor and which do
not. From this it follows that the resonator admits a range
of distinct operating configurations.
The treatment is theoretical throughout. Full-wave simulations of the
DALI proof-of-principle apparatus and the corresponding laboratory
measurements are collected in Appendices~\ref{app:sim}
and~\ref{app:lab}, and are used only to illustrate the results derived in
the main text.  Appendix~\ref{app:tm} establishes the relation of the
present description to the transfer-matrix formalism of a \textit{dielectric haloscope}~\cite{MADMAX:2019pub}.

\section{The instrument and its observable}\label{sec:instrument}

\begin{figure}[h]
\centering
\includegraphics[width=0.48\textwidth]{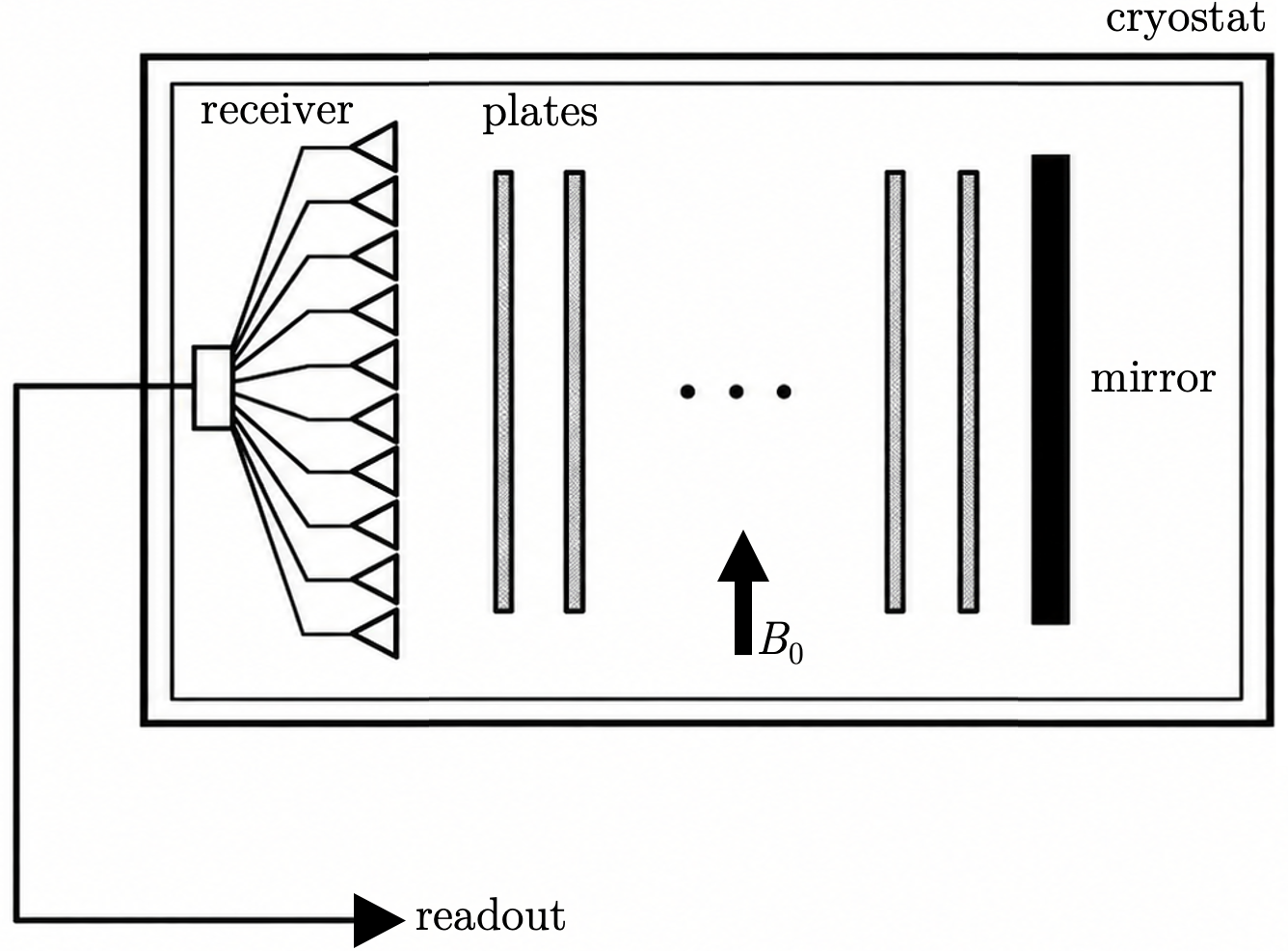}
\caption{Schematic view of a DALI haloscope. A stack of $N$ equally spaced layers is housed inside a vessel, terminated by a mirror at one end and coupled to a phased-array antenna receiver at the other. A superconducting solenoid (not shown) permeates the volume, providing a magnetic field $B_0$ parallel to the ceramic interfaces.}
\label{fig_0}
\end{figure}

The resonator consists of $N$ dielectric layers of relative permittivity
$\varepsilon_r$ and thickness $d_e$, stacked along $\hat{z}$ with vacuum
gaps $d_v$, backed by a conducting mirror, and immersed in a static
magnetic field $\bm{B}_0=B_0\hat{y}$. A single port at the opposite
termination couples the assembly to the receiver. A sechematic view of the haloscope is shown in Fig. \ref{fig_0}.

\begin{table}[b]
\caption{Orientation conventions used throughout.}
\label{tab:axes}
\begin{ruledtabular}
\begin{tabular}{ll}
Direction & Physical content \\
\colrule
$\hat{y}$ & $\bm{B}_0$, and hence the polarisation of $\bm{E}$ \\
$\hat{z}$ & Normal to the plates; propagation direction \\
$\hat{x}$ & The remaining transverse direction \\
\end{tabular}
\end{ruledtabular}
\end{table}

Because the plate normal is perpendicular to the magnetic field,
$\hat{n}_{\mathrm{plate}}=\hat{z}\perp\bm{B}_0$, every interface radiates
along $\pm\hat{z}$ with $\bm{E}\parallel\hat{y}$.  The longitudinal
problem is therefore one-dimensional in $z$, with the transverse
structure factorised.  Two areas appear and must be kept distinct:
$\Aphys$, the physical surface of the magnetized mirror, over which the
available power is defined; and $\Axy$, the transverse cross-section of
the magnetized volume, which enters the volume integrals of
Sec.~\ref{sec:longitudinal} through $dV=\Axy\,dz$.

The instrument has a single observable.  Whatever its origin, the signal
reaches the receiver as an electromagnetic standing wave between the
mirror and the port, and every quantity the experiment reports is
inferred from that wave.  This is not a simplification but a constraint
on what the analysis may legitimately use, and we return to it in
Sec.~\ref{sec:sourceindep}, where the wave is shown to be a property of
the resonator rather than of whatever excites it.

It is useful to recall at the outset the elementary properties of a
Fabry--P\'erot resonator in the limit at an isolated pole ~\cite{renk2012basics}. For a resonator of
optical length $L$ terminated by reflectors, the field follows from the
boundary conditions and the dispersion relation alone,
\begin{equation}
E_y(z,t)=A\,\sin(k_m z)\,\cos(\omega_m t),
\qquad k_m=\frac{m\pi}{L},
\label{eq:renkfield}
\end{equation}
with $m$ the order of the resonance.  Its sign alternates along $\hat z$,
while the stored energy density,
\begin{equation}
u(z)=\tfrac12\varepsilon_0\varepsilon_r A^{2}\sin^{2}(k_m z)\;\ge\;0 ,
\label{eq:renku}
\end{equation}
is positive definite everywhere.  Both statements hold in \emph{every}
Fabry--P\'erot resonator.  A sign-alternating longitudinal profile is
therefore the generic form of a Fabry--P\'erot eigenmode, and not a
feature of one stacking rather than another; the whole of
Sec.~\ref{sec:longitudinal} is an elaboration of the compatibility of
\eqref{eq:renkfield} with \eqref{eq:renku}.

Losses are collected in the survival factor $V$ per round trip, which
factorises into an out-coupling and an internal contribution,
$V=V_{\mathrm{out}}V_{\mathrm{i}}$.  With $T_{\mathrm{rt}}$ the
round-trip transit time, the photon lifetime, the effective number of
round trips and the quality factor are
\begin{equation}
\tau_p=\frac{T_{\mathrm{rt}}}{-\ln V},
\qquad
s_{\mathrm{eff}}=\frac{\tau_p}{T_{\mathrm{rt}}},
\qquad
Q=\omega_0\tau_p=2\pi\,m\,s_{\mathrm{eff}} .
\label{eq:renkQ}
\end{equation}
The factorisation of $V$ is the same statement as the separation
$1/\QL=1/Q_0+1/\Qext$, with $\beta=Q_0/\Qext$, used throughout and
collected in Appendix~\ref{app:oneport}: $V_{\mathrm{out}}$ carries the
coupling to the port and $V_{\mathrm{i}}$ the dissipation within the
stack.

Two features of \eqref{eq:renkfield}--\eqref{eq:renkQ} matter later.
Nothing in them refers to the source, so $Q$, $V$ and $\tau_p$ are
properties of the resonator and their measurement in one excitation
configuration transfers to any other.  And the derivation assumes a
quasiplane wave, for which the phase is constant over any plane
perpendicular to $\hat z$; the modes of a resonator of finite lateral
extent are diffractive and do not have this property, which is the single
respect in which the present treatment must go beyond the simplified
case.

\section{The model}\label{sec:model}

A result of this work is a closed expression for the power delivered at
the receiver,
\begin{equation}
\Pout=\underbrace{\frac{E_0^{2}}{2Z_0}\,\Aphys}_{P_0}
      \;\times\;\eta
      \;\times\;\QL
      \;\times\;\frac{\beta}{1+\beta} ,
\label{eq:chain}
\end{equation}
in which $E_0=\gagg B_0 a_0$ is the axion-induced field amplitude, $Z_0$
the impedance of free space, $\eta$ a dimensionless transverse
projection, and $\QL$ and $\beta$ the loaded quality factor and coupling
coefficient of the resonator. Each factor answers a separate question,
and the sections that follow establish them in turn.

The first factor, $P_0$, is the power available at the magnetized mirror.
Section~\ref{sec:P0} derives it and shows why it scales with the
conversion \emph{area} rather than with the magnetized \emph{volume}---a
point that is not obvious, since the axion field permeates the volume.

The factor $\QL$ is the resonant enhancement.  It is a measured quantity,
and Sec.~\ref{sec:sourceindep} establishes the property that makes its
measurement meaningful: the standing wave stored in the resonator does
not depend on what excites it, so a characterisation performed with an
external calibration source describes the instrument in its operating
configuration.  Section~\ref{sec:longitudinal} then shows that $\QL$, and
with it every quantity obtained by reflectometry, is blind to the
longitudinal phase structure of the mode, and reviews the observables
that establish resonant storage independently of that structure.

The factor $\eta$ is the transverse projection of the source onto the
mode read by the port.  Section~\ref{sec:transverse} defines it and shows
that it is the single coherent factor in Eq. \eqref{eq:chain}.

The last factor, $\beta/(1+\beta)$, is the extraction efficiency of a
one-port resonator, and follows from the definitions collected in
Appendix~\ref{app:oneport}.

It should be said at the outset which factor of Eq. \eqref{eq:chain} carries
what. The coherence with which the dielectric interfaces radiate acts on the power deposited in the mode per pass, and nowhere else; $\QL$
quantifies the resonant enhancement associated with the photon lifetime
and is blind to it. The number of recirculations is
$\mathcal{N}_{\mathrm{rt}}$ of Sec.~\ref{sec:longitudinal}, to which
$\QL=\omega_0\tau_p$ is proportional but which it does not equal.
Section~\ref{sec:sourceindep} establishes the separation and
Sec.~\ref{sec:family} draws its consequence for the design.  Neither
$\eta$ nor $\beta$ is contained in the other; $\eta$ belongs to the
source term and $\beta$ to the extraction term.  And the longitudinal
phase structure of the mode appears in none of the four factors: not in
$P_0$, which is a surface quantity; not in $\QL$ or $\beta$, which are
quadratic; and not in $\eta$, which is a purely transverse integral. The converse is equally true and is worth stating: a large $\QL$ does not
imply a large source projection.  The two are independent functionals of
the same eigenmode, and neither can be inferred from the other.

\section{The power available at the mirror}\label{sec:P0}

The axionlike wave is written
$a(t,\bm{x})=a_0\,e^{i(\bm{p}\cdot\bm{x}-\omega t)}$, with the physical
field its real part.  In the low-velocity limit the axion-induced
electric field has amplitude
\begin{equation}
E_0=\gagg B_0 a_0
   \simeq 1.3\times10^{-12}\,\frac{\mathrm{V}}{\mathrm{m}}
     \left(\frac{B_0}{10\,\mathrm{T}}\right)
     C_{a\gamma\gamma}\,f_{\mathrm{DM}}^{1/2},
\label{eq:E0}
\end{equation}
where $f_a(\bm{x})=f_{\mathrm{DM}}\rho_0$ is the distribution function of
axion dark matter across the vessel, $f_{\mathrm{DM}}$ accounts for
distributions other than a standard isotropic halo of local density
$\rho_0\simeq0.45\;\mathrm{GeV\,cm^{-3}}$~\cite{Staudt:2024tdq}, and
$C_{a\gamma\gamma}$ is the axion--photon coupling constant.

At a perfectly conducting surface the tangential field must vanish; this
is the dish-antenna principle~\cite{Horns:2012jf}.  The axion-induced
field $-E_0\hat{y}$ is therefore cancelled by an emitted wave of
amplitude $E_0$, whose cycle-averaged Poynting flux is $E_0^{2}/2Z_0$
with $Z_0=\sqrt{\mu_0/\varepsilon_0}$.  Integrating over the magnetized
mirror,
\begin{equation}
P_0=\frac{E_0^{2}}{2Z_0}\,\Aphys
 \sim 2\times10^{-27}\,\mathrm{W}\;
   \frac{\Aphys}{\mathrm{m}^2}
   \left(\frac{B_0}{10\,\mathrm{T}}\right)^{2}
   C_{a\gamma\gamma}^{2}\,f_{\mathrm{DM}}.
\label{eq:P0}
\end{equation}

\subsection{The conversion is a surface effect}

Equation~\eqref{eq:P0} scales with area, not with volume.  The axion
field permeates the whole magnetized region, and one might expect a
longer magnetized length to deliver proportionally more power.  It does
not, and the reason is already contained in the structure of the
axion-modified equations.

Within a homogeneous magnetized medium of permittivity $\varepsilon$ the
axion-induced field
\begin{equation}
\bm{E}_a=-\,\frac{\gagg\,a\,\bm{B}_0}{\varepsilon}
\label{eq:Ea}
\end{equation}
is uniform in space, and being uniform it is a particular solution of the
equations in the bulk: it radiates nothing.  A homogeneous solution---a
propagating wave---must be added only where the particular solution fails
to satisfy the interface conditions, that is, where $\varepsilon$ changes
discontinuously or where a conductor imposes $E_{\parallel}=0$.  Emission
therefore originates at surfaces and not in the volume between them, and
the conversion area rather than the magnetized volume is the quantity
that enters Eq. \eqref{eq:P0}.

A residual objection survives this argument and is worth settling
separately, since it is the one usually raised. If every volume element
of the effective current is regarded as an elementary radiator, why do
the elements deep inside the magnet not contribute?  Consider a source
distributed uniformly and in phase over $0\le z\le L$, radiating along
$\pm\hat{z}$, with a conducting mirror at $z=0$; this is the axion case
in the low-velocity regime, in which the de Broglie wavelength greatly
exceeds $L$ and the phase is common to the whole region.  An element at
$z'$ radiates in both directions.  The forward wave reaches the exit
plane $z=L$ with phase $e^{ik(L-z')}$; the backward wave travels to the
mirror, reflects with coefficient $-1$, and reaches the same plane with
phase $-e^{ik(z'+L)}$.  Their sum,
\begin{equation}
e^{ik(L-z')}-e^{ik(z'+L)}=-2i\,e^{ikL}\sin(kz') ,
\label{eq:pair}
\end{equation}
shows that the effective weight of the element at $z'$ is $\sin(kz')$:
the eigenmode of the mirror-terminated region.  The mirror reflection is
thus already contained in the modal weight and need not be added
separately.  Integrating over the source,
\begin{equation}
\int_0^{L}\sin(kz)\,dz=\frac{1-\cos kL}{k}\;\le\;\frac{2}{k}
=\frac{\lambda}{\pi} .
\label{eq:satur}
\end{equation}
The integral oscillates with $L$ and never exceeds $\lambda/\pi$, a bound
set by the wavelength alone.  Elements deeper than that contribute in
antiphase with elements nearer the exit plane and cancel against them, so
that extending the magnetized region cannot increase the coupling.  The
two arguments agree, as they must: the depth over which a homogeneous
magnetized volume radiates is fixed by the wavelength and not by the
extent of the magnet.

The bound is a statement about a uniform source integrated against a mode
of a single spatial frequency, and the essential ingredient is the
uniformity of the refractive index, not the presence of the mirror. The
same bound follows for $\int_0^L e^{-ikz}dz$ in the absence of a
terminating reflector.  Where $\varepsilon_r(z)$ varies, the mode is no
longer a single Fourier component and the bound does not apply, which is
the regime taken up in Sec.~\ref{sec:family}.

The same calculation has a second use.  Since the weight of an element at
$z'$ is $e_m(z')$, an extended source is equivalent, for the purpose of
exciting the mode, to a localised one.  Defining the effective coherent
length
\begin{equation}
\ell_{\mathrm{eff}}\equiv
\frac{\left|\int_0^{L}e_m(z)\,dz\right|}{\max_z|e_m(z)|} ,
\label{eq:leff}
\end{equation}
a distributed source of density $J_0$ excites the mode exactly as would a
localised source of strength $q=J_0\ell_{\mathrm{eff}}$ placed at an
antinode.  The reduction is exact, and $\ell_{\mathrm{eff}}$ is a
property of the mode alone: it is fixed by $\varepsilon_r(z)$ and the
boundary conditions, not by the source.  In the dielectric-free region
$e_m=\sin kz$ and Eq. \eqref{eq:satur} gives
$\ell_{\mathrm{eff}}\le\lambda/\pi$; with the stack present
$\ell_{\mathrm{eff}}$ follows from the loaded eigenmode and may exceed
that value.  This reduction is what allows the results of
Sec.~\ref{sec:sourceindep}, stated there for arbitrary sources, to be
applied to an axion source distributed through the magnetized volume.

\subsection{The emitting assembly}

In the multilayer assembly the total emission is the sum of the mirror
contribution and those of the $2N$ dielectric interfaces.  The latter are
governed by the coherence with which they radiate, quantified in
Sec.~\ref{sec:family}: they cancel among themselves in one limit and add
in the other, and the two limits are realised by different choices of
spacing.  In either case the emission of the magnetized mirror alone
provides the natural reference, and $P_0$ is adopted as such throughout.
It is also the normalisation against which \textit{boost factors} are
conventionally quoted, so that Eq. \eqref{eq:chain} and the boost-factor
formalism share a common baseline.

The choice has a practical consequence worth recording.  Configurations
that do not require coherent addition admit plate spacings substantially
smaller than those dictated by it, and the reduced longitudinal footprint
permits stacks of many layers to fit within the bores of conventional
superconducting solenoids---an advantage that grows as the target
frequency falls and the wavelength, and with it the coherent spacing,
increases.

\section{The observable standing wave does not depend on its source}
\label{sec:sourceindep}

The instrument is characterised by injecting a signal through the port
and measuring the reflected one.  In the operating configuration nothing
is injected: the resonator is populated from within, and the antenna
receives.  That the first configuration describes the second is not
self-evident, and it is established here.

\subsection{Coherence of the drive}

The axion field remains coherent for longer than the resonator
remembers, and this sets a design limit rather than being an idle
assumption. Its de Broglie wavelength largely exceeds the length of the assembly, so that its phase is uniform across the
resonator.  Its quality factor is $Q_a\sim10^{6}$; even for loaded
quality factors up to $\QL\sim10^{5}$ the coherence time exceeds the
photon lifetime by an order of magnitude and the axion line remains
inside the resonator bandwidth, so that the drive is monochromatic over
the interval in which the field accumulates.  Loaded quality factors
approaching $Q_a$ would place the resonator bandwidth below the axion
linewidth and forfeit part of the signal, so that $\QL\lesssim Q_a$ is
the ceiling of the resonant enhancement.

\subsection{The mode belongs to the resonator}

The eigenmodes $e_n(z)$ satisfy
\begin{equation}
\frac{d^{2}e_n}{dz^{2}}+\frac{\omega_n^{2}}{c^{2}}\,\varepsilon_r(z)\,e_n=0
\label{eq:helm}
\end{equation}
with the boundary conditions of the system, are orthogonal with weight
$\varepsilon_r$, and may be normalised so that
$\int\varepsilon_r e_m e_n\,dz=\delta_{mn}$.  Neither the differential
equation nor the normalisation contains any reference to a source.

Maxwell's equations with a source current $J_y$ give, for the $y$
component, $\partial_z^{2}E_y-(\varepsilon_r/c^{2})\partial_t^{2}E_y
=\mu_0\partial_t J_y$.  Expanding $E_y=\sum_n a_n(t)e_n(z)$, projecting,
and introducing loss phenomenologically through $Q$,
\begin{equation}
\ddot{a}_n+\frac{\omega_n}{Q}\,\dot{a}_n+\omega_n^{2}a_n
 =-\frac{1}{\varepsilon_0}\,\dot{\mathcal{F}}_n ,
\qquad
\mathcal{F}_n\equiv\int e_n J_y\,dz .
\label{eq:driven}
\end{equation}
The source enters only through the scalars $\mathcal{F}_n$.  It cannot
modify $e_n$, which solves the homogeneous problem.  No restriction has
been placed on the spatial support of $J_y$; a point source, a surface
source and a source distributed through the magnetized volume are all
particular cases.

Near an isolated resonance the $n=m$ denominator of Eq. \eqref{eq:driven} is
smaller than the others by a factor of order $Q$ times the fractional
mode spacing.  When that spacing exceeds the linewidth by one or two
orders of magnitude, the sum collapses to $E_y(z)\simeq a_m e_m(z)$, and
the spatial profile is $e_m(z)$ for \emph{any} source distribution.  A
sharper form of this condition, phrased in terms of the eigenvalues of
the round-trip operator of the stack, is given in
Appendix~\ref{app:tm}.

\subsection{The projection of a distributed source}

For a source distributed through the magnetized volume the projection in
Eq. \eqref{eq:driven} is the volume integral
$\mathcal{F}_m=\int e_m J_y\,dV$.  Because the axion-induced field
Eq. \eqref{eq:Ea} is uniform within each homogeneous component, that integral
may be rewritten, using the interface conditions, as a sum over the
emitting surfaces of the assembly,
\begin{equation}
\mathcal{F}^{\mathrm{ax}}_m
=\mathcal{F}_{\mathrm{mirror}}+\sum_{i=1}^{2N}s_i\,e_m(z_i),
\label{eq:Fsum}
\end{equation}
with weights $s_i\propto\Delta(1/\varepsilon)_i$ at the dielectric
interfaces.  The rewriting is an identity and introduces no assumption.

The two kinds of terms in Eq. \eqref{eq:Fsum} are of different character, by
construction rather than by oversight.  The terminating mirror enters the
eigenmode through the boundary condition, so that it carries no modal
weight of its own; $\mathcal{F}_{\mathrm{mirror}}$ is instead the
reference projection, defined as the one corresponding to the power $P_0$
of Sec.~\ref{sec:P0}, and it is adopted as such throughout.  The
interface terms, by contrast, are evaluated from the eigenmode itself.
Taking $\mathcal{F}^{\mathrm{ax}}_m=\mathcal{F}_{\mathrm{mirror}}$ is
therefore the statement that the assembly emits as its magnetized mirror
alone, and it is in that sense that Eq. \eqref{eq:chain} is referred to a
single mirror of equal area.

No emitting surface is omitted in Eq. \eqref{eq:Fsum}, and no assumption is
made about the relative signs of the terms: the results of the following
subsection hold whatever their values, since they affect
$\mathcal{F}^{\mathrm{ax}}_m$ and hence only an overall scale.

\subsection{Equal power, equal standing wave}

Solving Eq. \eqref{eq:driven} in steady state at $\omega=\omega_m$ gives
$a_m=-(Q/\omega_m\varepsilon_0)\mathcal{F}_m$, whence the stored energy
and the power the source must supply to sustain it are,
\begin{equation}
U=\frac{Q^{2}|\mathcal{F}_m|^{2}}{2\varepsilon_0\omega_m^{2}},
\qquad
\Pin=\frac{\omega_m U}{Q}=\frac{Q\,|\mathcal{F}_m|^{2}}{2\varepsilon_0\omega_m}
    =Q\,\Pmode ,
\label{eq:PinF}
\end{equation}
with $\Pmode=|\mathcal{F}_m|^{2}/(2\varepsilon_0\omega_m)$ the power the
same source would deliver in a single pass, in the absence of resonant
build-up.  In steady state the source acts on a field already enhanced by
the resonance, which is why $\Pin$ exceeds $\Pmode$ by the factor $Q$.

Equation~\eqref{eq:PinF} is a product of two factors that answer
different questions and are computed from different objects.  The
quantity $\Pmode$ is what the source deposits in the mode in a single
pass, and it is built from $\mathcal{F}_m$, a \emph{linear} functional of
the field weighted against the source distribution.  The factor $Q$ is
the number of times that deposit accumulates before it is lost, and it is
a ratio of two \emph{quadratic} functionals of the same field,
$\omega U/\Ploss$.  A quadratic functional is invariant under
$e_m\to-e_m$ over any subregion and cannot register the relative phases
with which different parts of an extended source contribute; a linear one
does nothing else.  The coherence of the interface emissions can
therefore act on $\Pmode$ and on nothing else, and $\QL$ cannot report it.

The contrast is sharpest when the two projections are written side by
side.  A localised source at the port and a source distributed through
the magnetized volume project onto the same eigenmode as
\begin{equation}
\begin{split}
\mathcal{F}_{\rm port}=\int e_m(z)\,\delta(z-z_p)\,dz=e_m(z_p) ,\\
\qquad
\mathcal{F}_{\rm ax}=J_0\!\int e_m(z)\,dz ,
\end{split}
\label{eq:twoproj}
\end{equation}

the first a single evaluation of the eigenmode and the second an integral
over it, in which contributions of opposite sign cancel against one
another.  Reflectometry depends on the mode through the first alone: a
measurement of $S_{11}$ and $\tau_g$ at one frequency returns two
numbers, and two numbers determine $\QL$ and $\beta$ and nothing further.
The coherence with which the interface emissions add is a third
functional of the same eigenmode, independent of those two, and no pair
of scalars can determine three independent quantities.

This limits what the measurement carries, not what is knowable.  The
profile $e_m(z)$ determines all three, and it may be obtained either from
a full-wave solution or from a field map of the resonator; the port
serves perfectly well to excite it, since by Eq. \eqref{eq:prop} the
eigenmode does not depend on the source that populates it.  What
reflectometry alone does not supply is the longitudinal integral of Eq.
\eqref{eq:twoproj}, so that $\QL$ and the coherence of the interface
emissions must be obtained from different measurements even though both
are properties of the same standing wave.

Writing the projection of Eq. \eqref{eq:Fsum} with its reference factored
out,
\begin{equation}
\mathcal{F}^{\mathrm{ax}}_m
=\mathcal{F}_{\mathrm{mirror}}
 \left[1+\frac{\sum_i s_i\,e_m(z_i)}{\mathcal{F}_{\mathrm{mirror}}}\right] ,
\label{eq:Fbracket}
\end{equation}
the bracket collects everything the ceramic interfaces contribute
relative to the mirror alone.  It approaches unity when the interface
contributions cancel among themselves, and departs from it when they add;
the parameter that measures this is introduced in
Sec.~\ref{sec:family}.  The delivered power then reads
\begin{equation}
\Pin=Q\;
\left|1+\frac{\sum_i s_i\,e_m(z_i)}{\mathcal{F}_{\mathrm{mirror}}}\right|^{2}
\;\frac{|\mathcal{F}_{\mathrm{mirror}}|^{2}}{2\varepsilon_0\omega_m} ,
\label{eq:Pinfact}
\end{equation}
a reference, a bracket, and a multiplicity, in that order.  The
enhancement $Q$ multiplies whatever the bracket returns; it does not
distinguish the mirror contribution from the interface contributions, and
it is not enlarged when they add coherently.

Throughout this work Eq. \eqref{eq:chain} is quoted with the bracket set to
unity, that is, per unit of source projection referred to the magnetized
mirror.  This is a choice of baseline rather than a claim about the
assembly, and it is the baseline against which boost factors are
conventionally reported~\cite{Millar:2016cjp}. A stacking whose interfaces add coherently
returns a bracket larger than unity and delivers correspondingly more
power; the present accounting does not claim that increment.  Whatever
its value, the bracket multiplies Eq. \eqref{eq:chain} as a whole and leaves
unchanged the three factors that follow it.  It is deliberately kept
outside Eq. \eqref{eq:chain} rather than written as a further factor: the
bracket is a signed longitudinal sum, whereas every factor of Eq.
\eqref{eq:chain} is either a functional of $|E|^{2}$ or a transverse
integral, and Sec.~\ref{sec:longitudinal} shows that no measurement of
the first kind can return a quantity of the second.  Writing the two in
one product would present a formula whose factors no single experimental
protocol supplies.

The source enters Eq. \eqref{eq:PinF} exclusively through $|\mathcal{F}_m|$.
Two sources that deliver the same power to the mode therefore store the
same energy and drive the same modal amplitude; and since the profile is
$e_m(z)$ in both cases, they produce the same standing wave, identical in
shape and in amplitude.  Equality of the complex $\mathcal{F}_m$ fixes in
addition the global phase.

More generally, and without any hypothesis on the delivered power, two
sources exciting the same isolated resonance produce fields that are
proportional,
\begin{equation}
E^{(1)}_y(z)=\chi\,E^{(2)}_y(z),
\qquad
\chi=\frac{\mathcal{F}^{(1)}_m}{\mathcal{F}^{(2)}_m}\in\mathbb{C},
\label{eq:prop}
\end{equation}
with $\chi$ independent of $z$, because $e_m$ cancels identically in the
ratio.  Every functional $\Phi[E]$ satisfying $\Phi[\chi E]=\Phi[E]$
therefore takes the same value for both. Note that $\chi$ is an arbitrary normalization fixed only by the local axion density; while the argument is more general and can be formulated independently of this overall scale. It is always possible to match the power induced by the axion field with that from an external source coupled to the system.

That class is broad, and it contains every quantity used to characterise
the resonator: the normalised profile $E_y/\max|E_y|$, the node
positions, $\QL$, $\beta$, $S_{11}$, $\tau_g$, the ratio of stored energy
between any two subregions, and the longitudinal coherence
$\int E_y\,dz/\!\int|E_y|\,dz$.  In this precise sense the standing wave
excited by a calibration source and that excited by the axion field are
the same standing wave, and the characterisation performed with the
former transfers in full to the latter.  The single quantity outside the
class is the overall scale $|\chi|$, which is set by the source coupling
and not by the resonator.

\section{The longitudinal phase structure does not enter $\QL$, $\beta$
or $\eta$}
\label{sec:longitudinal}

In a multilayer resonator the field alternates sign between successive
lobes along $\hat z$ for some choices of spacing and does not for others.
It is natural to ask whether that difference propagates into the
extracted power.  It does not, for a reason that is elementary once the
relevant functionals are separated.

\subsection{The measured quantities are quadratic}

The stored energy and the dissipated power are
\begin{equation}
\begin{aligned}
U &= \tfrac12\!\int\!\left(\varepsilon_0\varepsilon_r|E|^2
     + \mu_0|H|^2\right)\,dV, \\
\Ploss &= \tfrac12\omega\!\int\!
\varepsilon_0\varepsilon_r\tan\delta\,|E|^2\,dV ,
\end{aligned}
\end{equation}
with $dV=\Axy\,dz$ and $Q\equiv\omega U/\Ploss$.  Both integrands are
$|E|^{2}$; replacing $E\to-E$ over any subregion leaves them unchanged.
The reflection coefficient $S_{11}=b_1/a_1$ is a ratio of wave amplitudes
at the port reference plane, and $\beta$ follows from $|S_{11}|$ at
resonance; neither involves a signed integral of the field over the
resonator.  Squaring discards the spatial phase, and the measurement of
$Q$, $\beta$ and $S_{11}$ is therefore insensitive to the longitudinal
phase structure of the standing wave, including a $+-+-$ alternation.

Two properties follow. First, $Q$ is homogeneous of degree zero in the
field amplitude: $U$ and $\Ploss$ both scale as $|a_m|^{2}$, so their
ratio does not depend on how strongly the mode is excited.  Second, $Q$
is a property of the mode rather than of the source--mode coupling, and
may be defined from the complex eigenfrequency,
$Q=\mathrm{Re}(\omega)/2\,\mathrm{Im}(\omega)$.

\subsection{Two functionals of the same field}

The stored energy and the circulating and extracted power do not cancel;
they go as $|E|^{2}$ and lobes of opposite sign add positively.  This is
fully compatible with $\int E_y\,dz\approx0$, which is a different
functional of the same field.  The two are not related by squaring,
\begin{equation}
\left(\int E_y\,dz\right)^{2}\;\neq\;\int |E_y|^{2}\,dz ,
\label{eq:twofunc}
\end{equation}
as the elementary case $E_y=\cos kz$ over one period shows: the first
vanishes identically while the second equals $\pi/k$, and the field
itself vanishes only at isolated points.  A vanishing signed integral
implies neither a vanishing field nor a vanishing energy.  Both
propositions are simultaneously true, and they are simultaneously true in
every Fabry--P\'erot resonator, as \eqref{eq:renkfield} and
\eqref{eq:renku} already made plain.

\subsection{Establishing resonant storage}

Since the quantities that certify a resonance are quadratic, they can be
used to establish resonant storage without reference to the sign
structure of the mode.  Three such observables are available, and they
are mutually independent.

\paragraph{Absorbed power.}
An assembly terminated by a conducting mirror reflects essentially all
incident power away from resonance. In the single-pole limit, the absorbed fraction at resonance,
\begin{equation}
\frac{P_{\mathrm{abs}}}{P_{\mathrm{inc}}}=\frac{4\beta}{(1+\beta)^{2}} ,
\label{eq:absorbed}
\end{equation}
is obtained from $|\Gamma_0|$ alone through the relations of
Appendix~\ref{app:oneport}.  A dip approaching critical coupling
establishes that energy enters the assembly, irrespective of how the
interface emissions combine.

\paragraph{Dwell time.}
The round-trip transit time of the stack is
\begin{equation}
\tau_{\mathrm{rt}}
 =\frac{2\left[N d_e\sqrt{\varepsilon_r}+(N+1)\,d_v\right]}{c} ,
\label{eq:roundtrip}
\end{equation}
and the number of round trips executed before the field leaves the
assembly is $\mathcal{N}_{\mathrm{rt}}=\tau_p/\tau_{\mathrm{rt}}$, with
$\tau_p=\QL/\omega_0$ the photon lifetime.  A non-resonant reflection
returns $\mathcal{N}_{\mathrm{rt}}\simeq1$.  A dwell time exceeding
$\tau_{\mathrm{rt}}$ is direct evidence of stored energy and admits no
alternative interpretation.  That the group-delay maximum be coincident
in frequency with the $S_{11}$ minimum further discriminates a resonance
from a purely dissipative impedance mismatch, which would produce the dip
without the delay.

\paragraph{Scaling with layer count.}
For an assembly whose losses are dominated by the terminations rather
than by the dielectric bulk, the stored energy grows in proportion to the
number of series layers while the dissipated power remains fixed, giving
$\QL\propto N$; this is realised experimentally in
Appendix~\ref{app:lab}.  The scaling is a property of the resonator
alone: an instrumental artefact, or a feature of the enclosure rather
than of the stack, would not track the layer count.  Its persistence also
bounds the loss mechanism, since dielectric loss would cap $\QL$ at
$\sim1/\tan\delta$ rather than let it grow with $N$ to infinity.

\subsection{Extraction is immune to the sign alternation}

Energy is extracted through the aperture, the only opening of an
otherwise mirror-backed cavity.  The mode leaks through it at a rate
$\omega/\Qext$, and the extracted fraction of the stored energy is
$\beta/(1+\beta)$.

This is unaffected by the longitudinal sign structure for the reason
given above: the antenna samples the field over a transverse plane at a
fixed longitudinal position and does not integrate along $\hat z$.  Its
coupling is governed by the local modal amplitude at the aperture,
whereas a signed longitudinal integral is governed by the whole profile.
These are distinct functionals of the same eigenmode, related by no
inequality in either direction, and a configuration in which one is small
places no constraint on the other.

\section{Where the transverse structure enters}\label{sec:transverse}

The longitudinal structure of the mode has been disposed of.  The
transverse structure has not, and it enters the power budget once.

The fraction of the reference power $P_0$ that feeds the mode read by the
port is the coherent transverse projection of the uniform source onto the
reciprocal field of that port,
\begin{equation}
\eta=\frac{\left|\displaystyle\int_A \Erec(x,y)\,dA\right|^{2}}
          {\Aphys\displaystyle\int_A|\Erec(x,y)|^{2}\,dA},
\qquad
\Pmode=\eta\,P_0 ,
\label{eq:eta}
\end{equation}
where $\Erec$ is the complex transverse field of the mode that connects
to the receiving port.  Equation~\eqref{eq:eta} is a projection and not
an area; it is dimensionless, it equals unity for a flat transverse
profile, and it is bounded by $(0\leq\eta\leq )$ according to the Cauchy--Schwarz inequality. The factorisation Eq. \eqref{eq:eta} is not a convention but a condition on
the read-out.  The volume form factor conventionally quoted for a closed
cavity, $C=|\int E\,dV|^{2}/(V\!\int\varepsilon_r|E|^{2}dV)$, separates
into a transverse and a longitudinal factor---the first of which is
$\eta$---if and only if the reciprocal field is of the form
$f(x,y)\,g(z)$.  A port that samples a transverse plane at the
termination supplies such a field; a probe coupled to the local field
within the assembly does not, its reciprocal field being localised in all
three directions. In the latter case no separation into an area and a
longitudinal factor exists, and the volume integral must be evaluated as
it stands.  This is the structural difference between the present
accounting and that of a closed cavity read through a small aperture, and
the separability it requires is verified for the simulated modes in
Appendix~\ref{app:sim}.

Unlike $\QL$, $\beta$ and the $S_{11}$ dip, $\eta$ is \emph{not} a
functional of $|E_y|^{2}$. Transverse lobes of opposite phase subtract
in the amplitude $\int_A\Erec\,dA$, and $\eta$ is reduced accordingly.
It is the single place in Eq. \eqref{eq:chain} at which the transverse
structure of the mode enters, and it is also the single respect in which
the treatment departs from the textbook description recalled in
Sec.~\ref{sec:instrument}: for a quasiplane wave the transverse phase is
constant and $\eta=1$ identically, whereas the modes of a resonator of
finite lateral extent are diffractive and are not quasiplane.

Two boundaries of the definition should be stated. Equation
\eqref{eq:eta} accounts for the transverse projection and for nothing
else; it contains no longitudinal integral, the longitudinal behaviour of
the mode being the resonant build-up quantified by $\QL$.  And the
efficiency with which the antenna collects the emission is contained in
the measured $\beta$, not in $\eta$; the former belongs to the extraction
term and the latter to the source term, and they do not overlap.

Writing the single-pass power as $\eta P_0$ presumes that the modal
impedance seen at the mirror plane equals $Z_0$.  This is a design
condition rather than an approximation. A read-out port formed by an
oversized guide operated far above cutoff presents a wave impedance
$\Zport=Z_0[1-(\nu_c/\nu)^{2}]^{1/2}$, which approaches $Z_0$ as
$\nu\gg\nu_c$, so that the two impedances entering Eq. \eqref{eq:chain} and
Appendix~\ref{app:oneport} coincide by construction.

\section{A set of distinct tuners}\label{sec:family}

The results assembled in Eq. \eqref{eq:chain} hold for any stacking. It is
therefore natural to ask what the choice of stacking does control, and
the answer places the so-called dielectric haloscope~\cite{MADMAX:2019pub} at one
end of a range of configurations. Both ends of that range descend from the same tuner.  A magnetized
mirror radiating into free space~\cite{Horns:2012jf} delivers $P_0$ with
neither coherent addition nor storage, and is the common ancestor of what
follows.  Placing layers in front of it so that their emissions arrive in
phase multiplies $P_0$ by the square of their coherent sum; placing a
resonator in front of it multiplies $P_0$ by $\QL\beta/(1+\beta)$
instead. The two routes act on the same baseline through different
factors of Eq. \eqref{eq:chain}, and neither presupposes the other: a
stacking chosen for coherent addition is not thereby storing, and one
chosen for storage is not thereby failing to convert.

\subsection{The boost coherence}

Writing $\mathcal{F}_i$ for the contribution of the $i$-th interface to
the longitudinal projection onto the resonant mode, the coherent addition
of the interface emissions is governed by
\begin{equation}
\CB\;\equiv\;
\frac{\left|\sum_i \mathcal{F}_i\right|}{\sum_i \left|\mathcal{F}_i\right|}
\;\in\;[0,1] ,
\label{eq:boostcoh}
\end{equation}
which attains unity when all interfaces radiate in phase and vanishes
when consecutive contributions accumulate a relative phase of $\pi$.  It
is a condition on the \emph{positions} of the interfaces relative to the
phase of the eigenmode.

Resonant storage is a different condition.  It is governed by $\QL$ and
$\beta$, and is a statement about the \emph{reflectivity} of the assembly
and the impedance presented at the aperture.  As established in
Sec.~\ref{sec:sourceindep}, $\QL$ is a ratio of quadratic functionals and
cannot register the relative phases that Eq. \eqref{eq:boostcoh} measures.
What follows shows more: the two are controlled by different functions of
the geometry, so that a stacking may suppress one while preserving the
other.

Consider a single slab of relative permittivity $\varepsilon_r$ and phase
depth
\begin{equation}
\delta=\frac{2\pi\sqrt{\varepsilon_r}\,d_e}{\lambda_0} ,
\label{eq:phasedepth}
\end{equation}
embedded in vacuum.  Its amplitude reflectivity follows from summing the
internal reflections,
\begin{equation}
r(\delta)=\frac{r_{12}\left(1-e^{2i\delta}\right)}{1-r_{12}^{2}e^{2i\delta}} ,
\qquad
r_{12}=\frac{1-\sqrt{\varepsilon_r}}{1+\sqrt{\varepsilon_r}} ,
\label{eq:slabrefl}
\end{equation}
whence
\begin{equation}
|r(\delta)|=\frac{2\,|r_{12}|\,|\sin\delta|}
{\sqrt{1-2r_{12}^{2}\cos2\delta+r_{12}^{4}}} .
\label{eq:slabmod}
\end{equation}
Both $|\sin\delta|$ and $\cos2\delta$ are invariant under
$\delta\rightarrow\pi-\delta$, so that $|r|$ is \emph{exactly} symmetric
about $\delta=\pi/2$; it vanishes for a slab that is optically absent
(($\delta\to0$) and again at half-wave thickness ($\delta=\pi$), where its
two internal reflections cancel, and
is largest at quarter-wave thickness.

The coherence of the emission from the same slab behaves differently.
Its two faces carry weights of opposite sign, $\Delta(1/\varepsilon)$
being $1-\varepsilon_r^{-1}$ at the front face and
$\varepsilon_r^{-1}-1$ at the back, while the eigenmode advances in phase
by $\delta$ between them.  Their contribution to the projection is
therefore proportional to $1-e^{i\delta}$, and
\begin{equation}
\left|1-e^{i\delta}\right|=2\left|\sin\tfrac{\delta}{2}\right| ,
\label{eq:facecoh}
\end{equation}
which is monotonic on $0\le\delta\le\pi$: the faces cancel against each
other for a thin slab and add for a half-wave one.  The phase advance
between the two faces is that of a progressive wave, which is the reading
appropriate to an open resonator, whose eigenmode is complex; in the
closed lossless limit the ratio between the faces would be real and the
alternation would reduce to a change of sign, leaving unaffected the
monotonicity on which the comparison rests.

The two quantities are thus different functions of the same variable, one
symmetric about $\delta=\pi/2$ and the other monotonic across it.  Two
stackings placed at $\delta$ and at $\pi-\delta$ present slabs of
\emph{identical} reflectivity, and hence the same finesse and the same
capacity for resonant storage, while the coherence of their face emissions differs, since Eq. \eqref{eq:facecoh} is monotonic across $\delta = \frac{\pi}{2}$.
 Eq.
\eqref{eq:boostcoh} collects, in addition, the phases accumulated across
the vacuum gaps, but its origin is the per-slab behaviour of Eq.
\eqref{eq:facecoh}.

\subsection{The two limits}

Two steps carry Eq. \eqref{eq:Pinfact} to the delivered power. The reference
projection is the one that corresponds to $P_0$, so that
$|\mathcal{F}_{\mathrm{mirror}}|^{2}/2\varepsilon_0\omega_m=\eta P_0$ by Eq.
\eqref{eq:eta}; and the port extracts the fraction $\beta/(1+\beta)$ of
what the mode dissipates. Hence
\begin{equation}
\Pout=P_0\;\eta\;\QL\;\frac{\beta}{1+\beta}\;
\left|1+\frac{\sum_i s_i\,e_m(z_i)}{\mathcal{F}_{\mathrm{mirror}}}\right|^{2} ,
\label{eq:chainfull}
\end{equation}
of which Eq. \eqref{eq:chain} is the case in which the bracket is unity.  Two of these factors can be made large, and each of the two is the design
principle of a different device.

When $\CB\to1$ the interface emissions add, the bracket grows, and the
assembly behaves as a dielectric haloscope~\cite{MADMAX:2019pub}: the
figure of merit is the boost factor, the spacings are dictated by the
requirement of coherent addition, and the achievable bandwidth is bounded
by a sum rule on the boost amplitude~\cite{Millar:2016cjp}. This is the
half-wave configuration, $\delta\to\pi$, in which Eq. \eqref{eq:facecoh} is
maximal.

When $\CB\to0$ the interface contributions cancel among themselves and
the bracket returns to unity. The assembly emits as its magnetized
mirror alone, and the stack contributes not by radiating but by storing:
the power available at the mirror is recirculated
$\mathcal{N}_{\mathrm{rt}}$ times and delivered to the port with
efficiency $\beta/(1+\beta)$. Nothing in Eq. \eqref{eq:chainfull} fails in
this limit; the bracket is simply spent, and the remaining factors carry
the whole of the enhancement.

How this limit is reached matters, and the two available routes are not
equivalent. Cancellation obtained by making the layers optically thin,
$\delta\to0$, is degenerate: by Eq. \eqref{eq:slabmod} the layers then cease
to reflect, the assembly stores nothing, and the stack is optically
absent.  The useful route is cancellation obtained by \emph{position},
with layers that individually remain reflective---$\delta$ of order
$\pi/2$, where Eq. \eqref{eq:slabmod} is largest---arranged so that their
contributions to Eq. \eqref{eq:boostcoh} oppose one another.  It is in this
sense that $\CB\to0$ is compatible with, and indeed favourable to,
resonant storage.

The two enhancements are not alternatives imposed by a single parameter.
The reflectivity that sets $\QL$ is symmetric about $\delta=\pi/2$ while
the face coherence that sets the bracket is monotonic across it, so that
the two are governed by different functions of the same geometry. What Eq.
\eqref{eq:chainfull} adds is that they enter the delivered power as
independent factors: a boost factor bounds neither $\QL$ nor $\beta$, and
$\QL$ bounds neither the bracket nor the coherence it measures.  Neither
figure of merit is a measurement of the other.

They differ, however, in what an experiment returns. The factors $\eta$,
$\QL$ and $\beta$ follow from the transverse structure of the mode and
from one-port reflectometry. The bracket does not: it is a signed
longitudinal sum, and Sec.~\ref{sec:longitudinal} shows that no
functional of $|E|^{2}$ can return one. It is for that reason, and not
because the bracket is absent or negligible, that \eqref{eq:chain} is
quoted with it at unity throughout this work; a device operated towards
$\CB\to1$ would deliver correspondingly more, and establishing how much
requires a determination of a different kind.

Both limits are operating points of the same assembly, and the
Fabry--P\'erot resonator admits all the intermediate configurations. The
choice among them is governed by the accessible frequency band, by the
room available for plate stacking, as phase coherence requires for a half-wavelength configuration which is more demandanding in terms of layer spacing, by the mechanical tolerances each
demands, and by the transverse projection $\eta$ of
Sec.~\ref{sec:transverse}. The two limits are idealisations; a real device may lie anywhere between them, and its exact position is case-dependent and better addressed through simulation and measurement than through purely theoretical derivation.


\section{Conclusions}\label{sec:conclusions}

We have derived the power delivered at the receiver of a semi-open multilayer
Fabry--P\'erot haloscope and identified what each factor of that
expression depends upon.  The power available at the magnetized mirror is
a surface quantity, and we have shown why an extended in-phase source
saturates rather than accumulating with the magnetized length.  The
resonant enhancement and the extraction efficiency are properties of the
resonator and its port, transferable from a calibration measurement to
the operating configuration by standard reciprocity and by the independence of the
eigenmode from whatever excites it.

The longitudinal phase structure of the mode enters none of these
factors.  The quantities that would carry it are functionals of
$|E|^{2}$, and squaring discards the sign; a sign-alternating profile is
in any case the generic form of a Fabry--P\'erot eigenmode.  The
transverse structure does enter, once, through a coherent projection
which we have defined and bounded.

Three consequences follow.  First, resonant storage and coherent addition
act on different factors of the delivered power: the first on the number
of recirculations, the second on the deposit per pass. They are
controlled by different functions of the geometry---the reflectivity of a
layer is symmetric about quarter-wave thickness while the coherence of
its faces is monotonic across it---so that a stacking may suppress the
second while preserving the first. The Fabry--P\'erot haloscope is therefore the less constrained device. It operates over the whole range of $\CB$, whereas a dielectric haloscope operating in the half-wave configuration has $\CB\simeq1$ by construction. That configuration buys a useful boost factor but pays for it with a constraint: the spacings are no longer free, being fixed by the requirement of coherent addition, more magnetized space in needed for plate stacking, and the tolerances on them tighten as the number of layers grows. A resonator operated at $\CB\to0$ gives up the boost and recovers the freedom.  Its spacings are set only by the requirement that the assembly resonate at the target frequency, a far weaker condition, and its amplification comes from a mechanism---resonant storage---that is measurable by simple one-port reflectometry and that scales with the number of layers as long as the losses remain termination-dominated.

Second, the axion field and a localised source such as a calibration
antenna generate the same observable: an electromagnetic standing wave
between the mirror and the port, identical in shape and phase and, for an arbitrary delivered power, in amplitude. This is what justifies using
reflectometric calibration and port-driven simulation to characterise the
instrument; it is not necessary to implement or to simulate a spatially
distributed axionlike source in order to establish $\QL$, $\beta$, or
the shape of the eigenmode. Neither is it necessary to perform internal measurements of the ceramic stack to determine the quality factor, since the external-port measurement determines it with the same accuracy.

Third, the description adopted here and the transfer-matrix description
conventionally applied to multilayer stacks are not competing accounts of
the same system.  Appendix~\ref{app:tm} shows that the factorisation into
$\mathcal{F}_m$ and $\QL$ is the isolated-resonance limit of the
transfer-matrix resummation, obtained by spectral decomposition of the
round-trip operator, and states the condition under which the two agree.
The transfer-matrix description remains valid where no single mode
dominates; the factorised one buys, with that assumption, the ability to
obtain each factor independently by one-port reflectometry.

The experimental case for resonant storage does not rest on any of the
foregoing.  A reflection minimum coincident in frequency with a
group-delay maximum, and a loaded quality factor growing linearly with
the layer count to $N\simeq20$ for even and odd $N$ alike, are
independent signatures, and neither a dissipative impedance mismatch nor
a feature of the enclosure would reproduce them.  They are collected in
Appendix~\ref{app:lab}.

\vspace{6pt}
\textit{\bf{Acknowledgements.}} The project that gave rise to these results received the support of a fellowship from “la Caixa” Foundation (ID 100010434). The fellowship code is LCF/BQ/PI24/12040023. J.DM. acknowledges support from the Spanish Ministry of Science, Innovation and Universities and the Agency (EUR2024-153552 financed by MICIU/AEI/10.13039/501100011033). We gratefully acknowledge financial support from the Severo Ochoa Program for Technological Projects and Major Surveys 2020-2023 under Grant No. CEX2019-000920-S; Recovery, Transformation and Resiliency Plan of Spanish Government under Grant No. C17.I02.CIENCIA.P5; Operational Program of the European Regional Development Fund (ERDF) under Grant No. EQC2019-006548-P; IAC Plan de Actuación 2022. This work is part of grant CEX2025-001609-S, awarded to the Instituto de Astrofísica de Canarias under the Severo Ochoa Centre of Excellence program and funded by MICIU/AEI/10.13039/501100011033. J.DM. acknowledges the support from the European Union via the UNDARK project of the Widening participation and spreading excellence programme (project number 101159929). M.M. acknowledges funding by the Deutsche Forschungsgemeinschaft (DFG, German Research Foundation) under Germany’s Excellence Strategy---EXC 2121 "Quantum Universe"---390833306. M.E.T. was funded by the Australian Research Council Centre of Excellence for Dark Matter Particle Physics (CE200100008).

\appendix

\section{One-port relations}\label{app:oneport}

For a single-port resonator the reflection coefficient near resonance is
\begin{equation}
\Gamma(\omega)=\frac{\beta-1-2jQ_0\delta}{\beta+1+2jQ_0\delta},
\qquad \delta=\frac{\omega-\omega_0}{\omega_0},
\label{eq:gamma}
\end{equation}
with $\beta=Q_0/\Qext$ and $1/\QL=1/Q_0+1/\Qext$.  At resonance
$|\Gamma_0|=|1-\beta|/(1+\beta)$ and the power balance reads
$P_{\mathrm{refl}}=|\Gamma_0|^{2}P_{\mathrm{inc}}$ and
$P_{\mathrm{abs}}=4\beta(1+\beta)^{-2}P_{\mathrm{inc}}$.

The magnitude $|\Gamma_0|$ does not determine $\beta$ on its own.  Both
\begin{equation}
\beta=\frac{1-|\Gamma_0|}{1+|\Gamma_0|}
\quad\text{and}\quad
\beta=\frac{1+|\Gamma_0|}{1-|\Gamma_0|}
\label{eq:branches}
\end{equation}
reproduce the same dip depth, the first under- and the second
overcoupled, and the two differ by $\beta\to1/\beta$.  The ambiguity is
resolved by the phase: the locus of $\Gamma(\omega)$ in the complex plane
is a circle, and the origin lies inside it if and only if the resonator
is overcoupled.  Equivalently, the sign of the group delay at resonance
distinguishes the two branches.  A value of $\beta$ quoted from $|S_{11}|$
alone is therefore incomplete, and the branch must be stated.  The loaded
quality factor is unaffected: under $\beta\to1/\beta$ one has
$Q_0\to Q_0/\beta$, and $\QL=Q_0/(1+\beta)$ is invariant.

Differentiating the phase of \eqref{eq:gamma} at $\delta=0$ gives the
group delay of the reflection measured in full at the port,
\begin{equation}
|\tau_g(\omega_0)|=\frac{4\,\QL\,\beta}{\omega_0\,|1-\beta|} .
\label{eq:taug}
\end{equation}
Equations~\eqref{eq:branches} and \eqref{eq:taug} describe an isolated
single pole, for which the reflection is exactly Lorentzian.  They are
stated here as idealisations, and are not the route by which the
quantities reported in this work are extracted in practice; the reason is given at
the end of this appendix.

The $\beta$ dependence of \eqref{eq:taug} originates in interference and
not in storage.  Writing \eqref{eq:gamma} as
\begin{equation}
\Gamma(\omega)=-1+\frac{1+\Gamma_0}{1+2j\QL\delta},
\qquad 1+\Gamma_0=\frac{2\beta}{1+\beta},
\label{eq:gammasplit}
\end{equation}
separates a frequency-independent direct term---the portion of the
incident wave that returns without entering the resonator---from the
resonant pole.  The first carries neither delay nor information on
$\QL$; the second carries both, and its residue $1+\Gamma_0$ is real and
therefore contributes nothing to the phase.  Where the direct term is
removed, the phase derivative is that of the pole alone,
\begin{equation}
\tau_g(\omega_0)=\frac{2\QL}{\omega_0},
\label{eq:taugated}
\end{equation}
independent of $\beta$.  The factor two is that between the decay of the
field amplitude, which a group delay measures, and the decay of the
stored energy, which defines $\QL$ through \eqref{eq:renkQ}.

Neither Eq. \eqref{eq:taug} nor Eq. \eqref{eq:taugated} is applied to the data of
Appendices~\ref{app:sim} and \ref{app:lab}.  An assembly of $N$ layers
supports several resonances within a band, and the observed lineshape is
not that of a perfect single pole: neighbouring poles contribute to the phase
derivative, and the coefficient relating $\tau_g$ to $\QL$ is then
neither of the two above.  In the simulations in Appendix~\ref{app:sim}, $\QL$ and $\beta$ are
obtained from a fit to the complex $S_{11}$ locus over a restricted
window, in which the direct term is the displacement of the circle from
the origin and is accounted for by the fit itself.  In Appendix~\ref{app:lab} the laboratory
response is transformed to the time domain, referred to that of a bare
mirror and filtered with a discrete prolate spheroidal window before the
delay is evaluated~\cite{PhysRevD.110.072013}, which removes the direct
term.  In both cases the group delay is used as a diagnostic---its
maximum must coincide in frequency with the $S_{11}$ minimum---rather
than as the source of $\QL$.  The photon lifetime $\tau_p=\QL/\omega_0$,
and with it $\mathcal{N}_{\rm rt}=\tau_p/\tau_{\rm rt}$, is accordingly
taken from the lineshape fit or from the complex eigenfrequency, and is
free of the branch ambiguity of Eq. \eqref{eq:branches}.

\section{Full-wave simulations}\label{app:sim}

The results of the main text are theoretical and none of them depends on
a particular geometry.  This appendix collects three-dimensional
finite-element simulations of the DALI proof-of-principle resonator that
illustrate them, exhibiting on one piece of hardware the two limits of
the family described in Sec.~\ref{sec:family}.

\subsection{The simulated assembly}

Geometry is in accordance with Table~\ref{tab:axes}. The model comprises $N=4$ dielectric layers of thickness $d_e=1$~mm and
relative permittivity $\varepsilon_r\sim25$, of lateral extent
$100\times100$~mm$^2$, separated by vacuum gaps $d_v=6.21$~mm and backed
by a conducting mirror at $z=35.05$~mm.  A single waveguide port at $z=0$
provides the excitation.  The fields were computed on a three-dimensional
mesh of several million elements and exported over the complete resonator
volume on a uniform $0.5$~mm grid.

The same hardware supports resonances at very different frequencies, and
the character of the stacking changes with them.  At $6.832$~GHz the
free-space wavelength is $\lambda=43.88$~mm, so that $d_v=\lambda/7.07$
and the optical thickness of a plate is
$d_e\sqrt{\varepsilon_r}=\lambda/8.78$; at $24.508$~GHz,
$\lambda=12.23$~mm and the same dimensions become $d_v=\lambda/1.97$ and
$d_e\sqrt{\varepsilon_r}=\lambda/2.45$.  We refer to the two as the
$\lambda/8$ and $\lambda/2$ configurations.  In terms of the phase depth
of Eq. \eqref{eq:phasedepth} they correspond to $\delta=0.23\pi$ and
$\delta=0.82\pi$; neither coincides exactly with the quarter-wave or
half-wave points, and the labels are used for brevity.  The round-trip
transit time, Eq. \eqref{eq:roundtrip}, is $\tau_{\rm rt}=0.341$~ns in both
cases.

The simulations are performed with $N=4$ for economy of computation.
Nothing in the derivations of Secs.~\ref{sec:P0}--\ref{sec:family} refers
to the layer count: $N$ enters only as the number of terms in the
interface sums of Eq. \eqref{eq:Fsum} and Eq. \eqref{eq:boostcoh} and through the
transit time of Eq. \eqref{eq:roundtrip}, so the conclusions hold for
arbitrary $N$.  The measured $\QL(N)$ behaviour of
Appendix~\ref{app:lab} validates the extension of the resonant-storage
mechanism from the four-layer simulation to stacks of at least twenty
layers.  It does not by itself fix the scaling of the source projection
$\mathcal{F}^{\mathrm{ax}}_m$, of $\CB$ or of $\eta$: these remain
independent quantities and must in general be evaluated separately.

\subsection{Reflectometry}
The simulator returns the reflection in full, with the direct term of Eq.
\eqref{eq:gammasplit} present, so that Eq. \eqref{eq:taug} governs the group
delay and its sign carries the coupling branch.  This is the opposite
situation to the laboratory measurement of Appendix~\ref{app:lab}, in
which that term is removed before the delay is evaluated.  In what
follows $\beta$ and its branch are taken from the locus of $\Gamma$ in
the complex plane and $\QL$ from a fit to the lineshape over a window of
a few linewidths; neither is obtained from $\tau_g$, which serves only as
the diagnostic of Sec.~\ref{sec:longitudinal}.
\begin{figure}[t]
\includegraphics[width=\columnwidth]{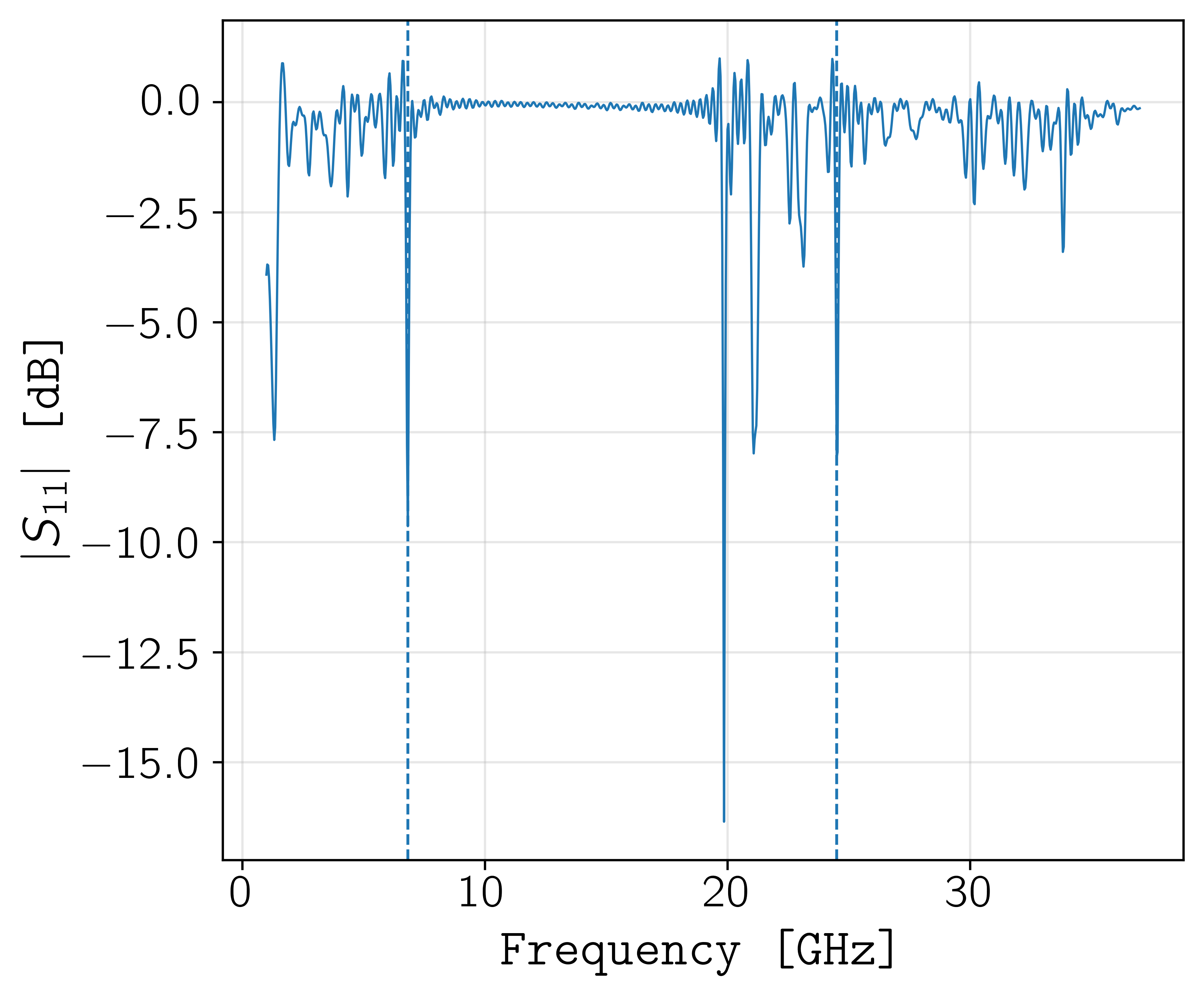}
\caption{Simulated reflection coefficient of the four-layer assembly over
the $1$--$37$~GHz band.  The two resonances used throughout this appendix
are marked: the $\lambda/8$ configuration at $6.832$~GHz and the
$\lambda/2$ configuration at $24.508$~GHz.  The same physical stack
supports both; what distinguishes them is the ratio of the layer
dimensions to the wavelength, and with it the relative phase of the
interface contributions.}
\label{fig:s11}
\end{figure}

\begin{figure}[t]
\includegraphics[width=\columnwidth]{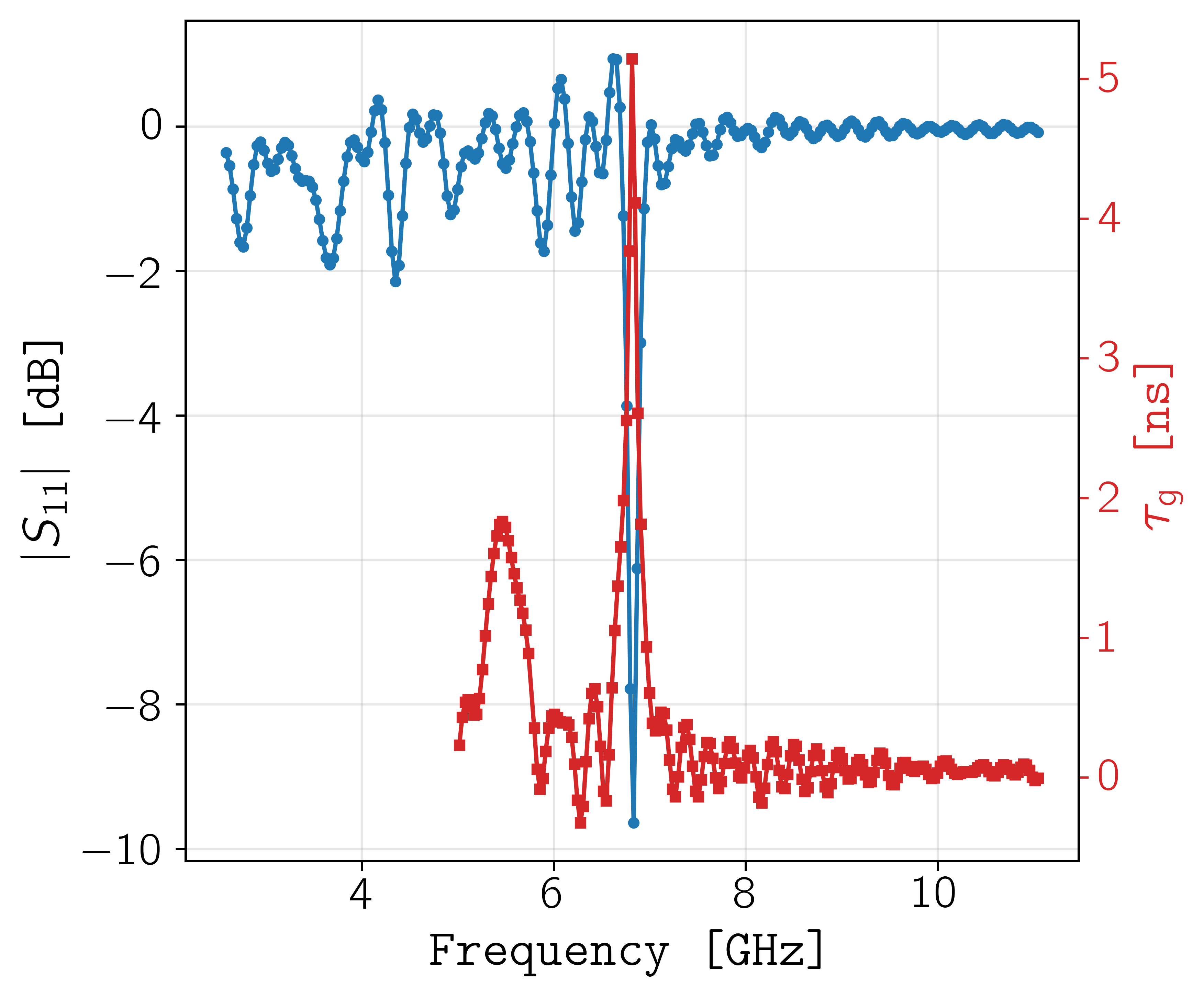}
\caption{Reflection coefficient (left axis) and group delay $\tau_g$
(right axis) in the neighbourhood of the $\lambda/8$ resonance.  At the
minimum, $|S_{11}|=-9.64$~dB, corresponding to $|\Gamma_0|=0.330$ and an
absorbed fraction $4\beta(1+\beta)^{-2}=0.891$. The locus of $\Gamma$
encircles the origin, which places the resonance in the overcoupled
branch of Eq. \eqref{eq:branches}, and the dip depth then gives
$\beta=1.98$; the group delay is positive at resonance, as Eq.
\eqref{eq:taug} requires in that branch.  The diameter of the fitted
circle, an estimator that uses the whole trace rather than the minimum
alone, returns $\beta=2.35$; the difference of eighteen per cent bounds
the departure of this resonance from a single isolated pole.  The two
quantities plotted are those a one-port measurement returns directly,
and their maxima coincide in frequency, which distinguishes a resonance
from a purely dissipative mismatch.}
\label{fig:s11l8}
\end{figure}

\begin{figure}[t]
\includegraphics[width=\columnwidth]{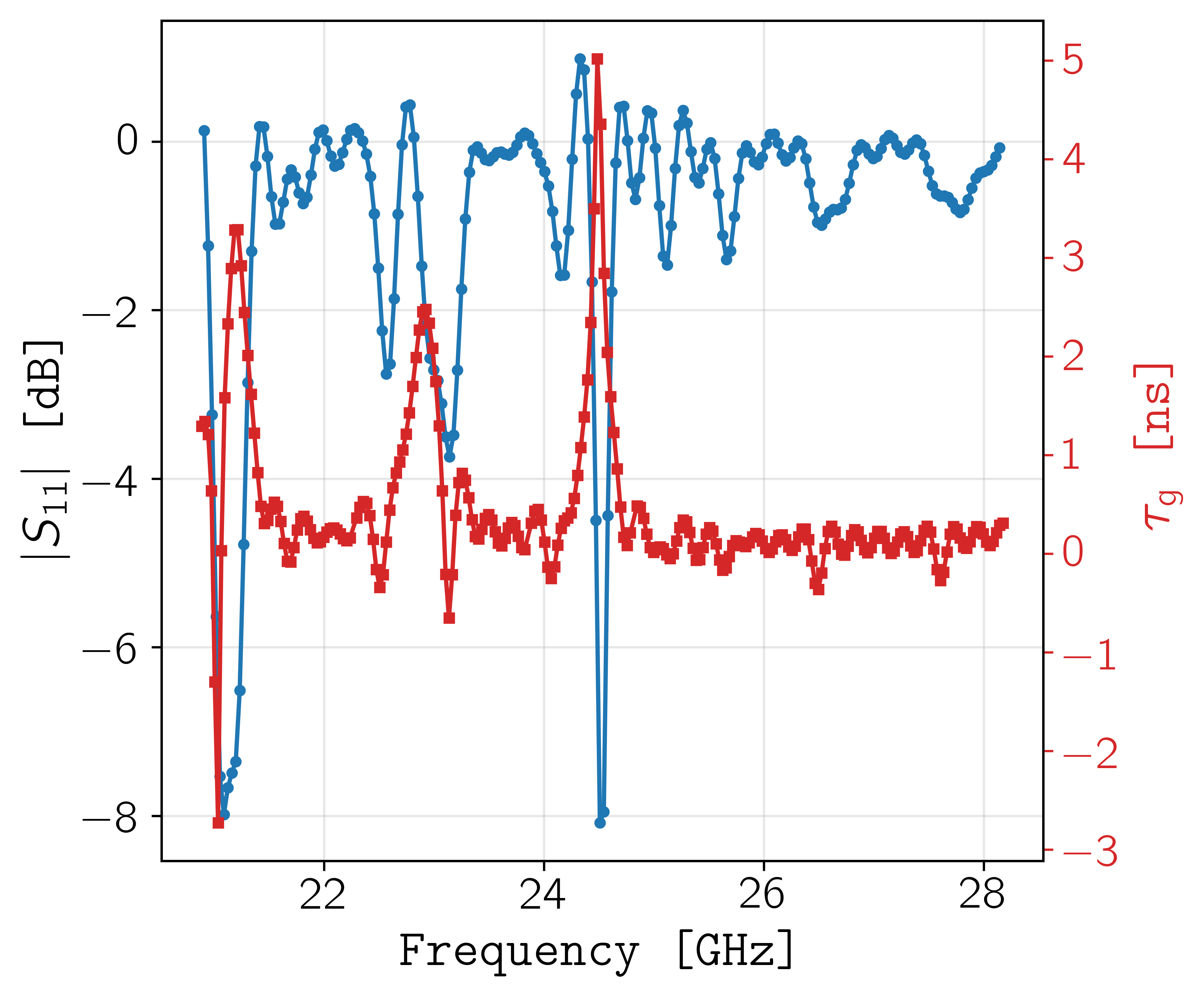}
\caption{As Fig.~\ref{fig:s11l8}, for the $\lambda/2$ resonance.  Here
$|S_{11}|=-8.08$~dB, $|\Gamma_0|=0.394$, $\beta=2.30$ and the absorbed
fraction is $0.844$; the circle estimator returns $2.48$, a closer
agreement than at $\lambda/8$.  The two configurations couple to the port
with comparable efficiency, $\beta/(1+\beta)$ differing by five per cent,
and their photon lifetimes agree to within about fifteen per cent,
although---as Appendix~\ref{app:cb} shows---they differ by more than an order
of magnitude in the coherence with which their interfaces radiate.}
\label{fig:s11l2}
\end{figure}

\subsection{Longitudinal profile and phase coherence}\label{app:cb}

\begin{figure}[h]
\includegraphics[width=\columnwidth]{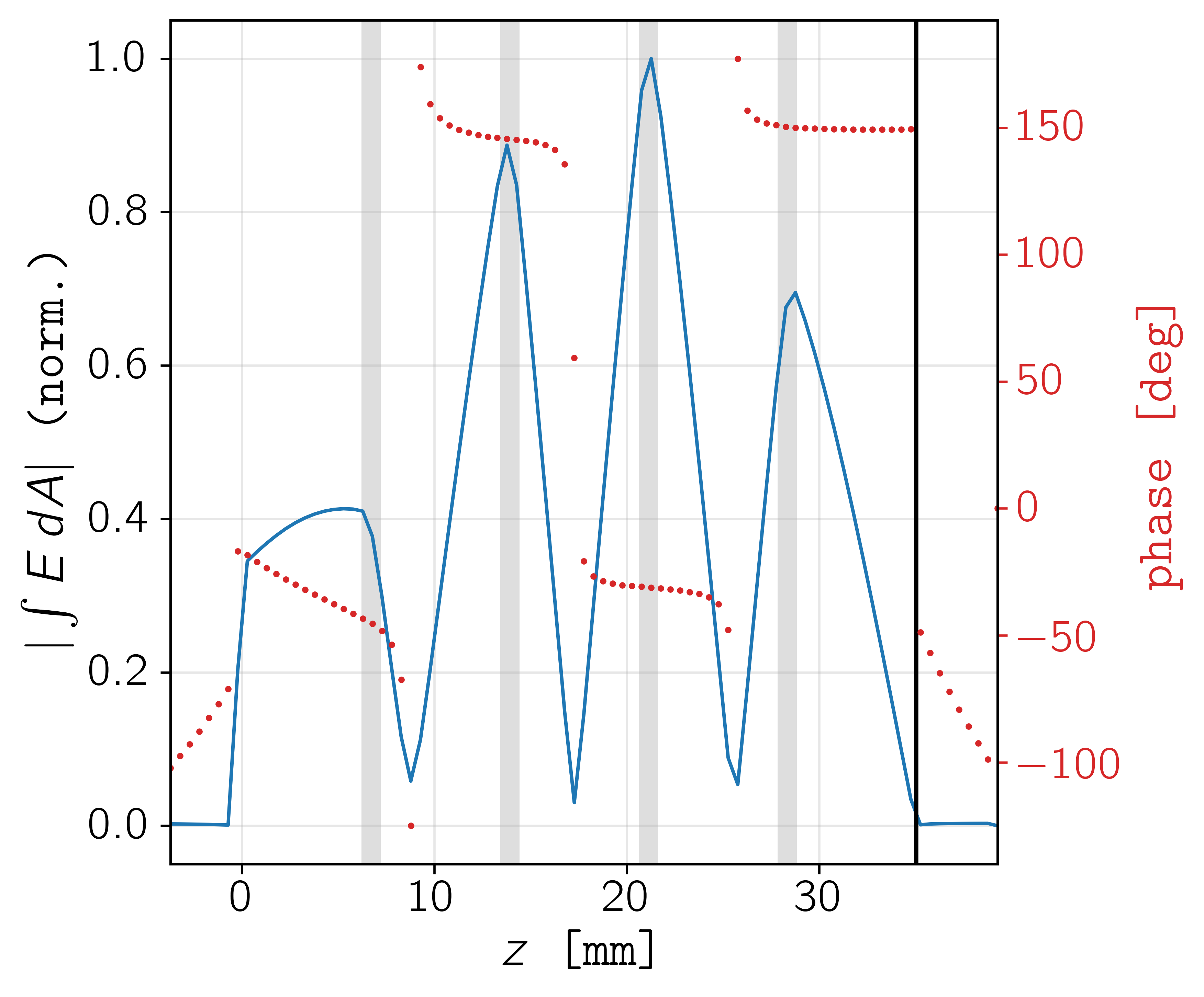}
\caption{Longitudinal profile of the mode in the $\lambda/8$
configuration.  The transverse integral $|\int E_y\,\mathrm{d}A|$,
normalised to its maximum, is shown against the left axis and its phase
against the right; the dielectric layers are shaded and the mirror plane
is the solid vertical line.  The phase takes two values separated by
$180^{\circ}$ and alternates between them at every node, which is the
signature of a standing wave with sign-alternating lobes and the generic
form of a Fabry--P\'erot eigenmode, Eq. \eqref{eq:renkfield}.  The
longitudinal coherence
$|\int E_y\,\mathrm{d}V|/\!\int|E_y|\,\mathrm{d}V$ is $1.20\%$.}
\label{fig:longl8}
\end{figure}

\begin{figure}[t]
\includegraphics[width=\columnwidth]{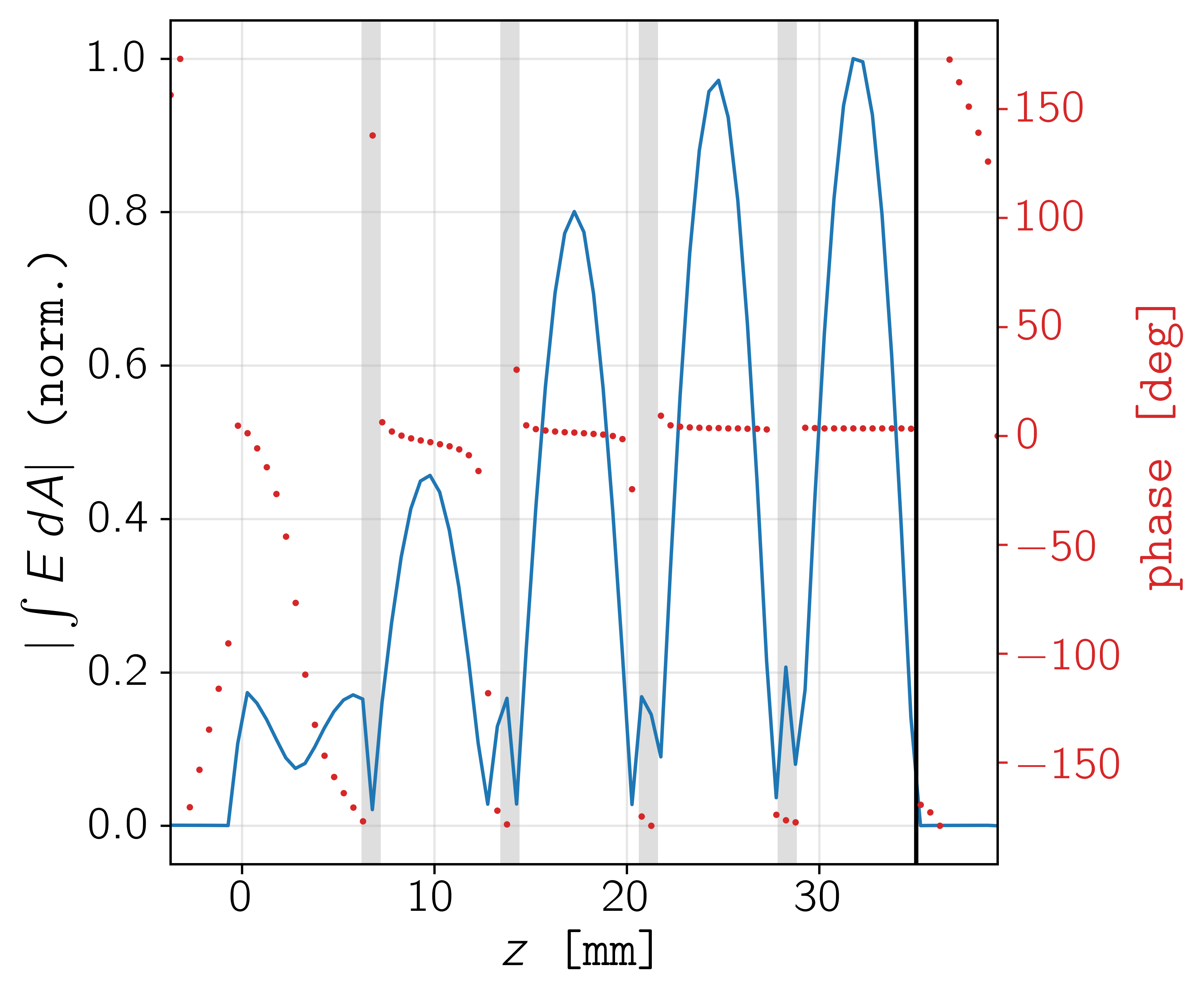}
\caption{As Fig.~\ref{fig:longl8}, for the $\lambda/2$ configuration.
The phase is now essentially constant across the whole assembly, with
brief excursions confined to the interiors of the dielectric layers, and
the longitudinal coherence rises to $75.9\%$.  The contrast with
Fig.~\ref{fig:longl8} is the entire content of the boost coherence $\CB$
of \eqref{eq:boostcoh}.}
\label{fig:longl2}
\end{figure}

The interface contributions to the longitudinal projection, and the
coherence with which they add, are collected in Table~\ref{tab:cb}.  In
the $\lambda/8$ configuration the eight phases alternate about a
$180^{\circ}$ separation and the sum retains $5.0\%$ of the incoherent
total; in the $\lambda/2$ configuration seven of the eight lie within
$30^{\circ}$ of one another and the sum retains $81\%$.  The two limits
of Sec.~\ref{sec:family} are therefore realised by the same stack at two
of its resonances.

\begin{table}[t]
\caption{Phase of the contribution of each dielectric interface to the
longitudinal projection onto the resonant mode, and the resulting boost
coherence $\CB$ of Eq. \eqref{eq:boostcoh}.  Plates are numbered from the
port; ``front'' and ``back'' denote the two faces of each.}
\label{tab:cb}
\begin{ruledtabular}
\begin{tabular}{lrr}
Interface & $\lambda/8$ [deg] & $\lambda/2$ [deg] \\
\colrule
P1 front & $-43.3$  & $-177.0$ \\
P1 back  & $+131.7$ & $-173.7$ \\
P2 front & $+146.0$ & $-172.0$ \\
P2 back  & $-34.8$  & $-149.6$ \\
P3 front & $-30.8$  & $-174.8$ \\
P3 back  & $+148.5$ & $-170.8$ \\
P4 front & $+151.0$ & $-173.9$ \\
P4 back  & $-30.0$  & $+2.6$   \\
\colrule
$\CB$ & $0.050$ & $0.810$ \\
\end{tabular}
\end{ruledtabular}
\end{table}

The effective coherent length of \eqref{eq:leff} makes the same point in
the language of Sec.~\ref{sec:P0}.  In the $\lambda/8$ configuration
$\ell_{\rm eff}=0.35$~mm, well below the dielectric-free saturation bound
$\lambda/\pi=13.97$~mm; in the $\lambda/2$ configuration
$\ell_{\rm eff}=11.94$~mm, above the corresponding bound
$\lambda/\pi=3.89$~mm.  The bound of Sec.~\ref{sec:P0} holds for a
region containing no dielectrics, and exceeding it is a qualitative
signature that the interfaces are radiating coherently.  The ratio
$\ell_{\rm eff}\pi/\lambda$ is not, however, a substitute for $\CB$: it is
normalised to the peak of the loaded mode rather than to the reference
projection of Eq. \eqref{eq:Fbracket}, and the two normalisations do not
cancel.

\subsection{Transverse projection}

\begin{figure}[h]
\includegraphics[width=\columnwidth]{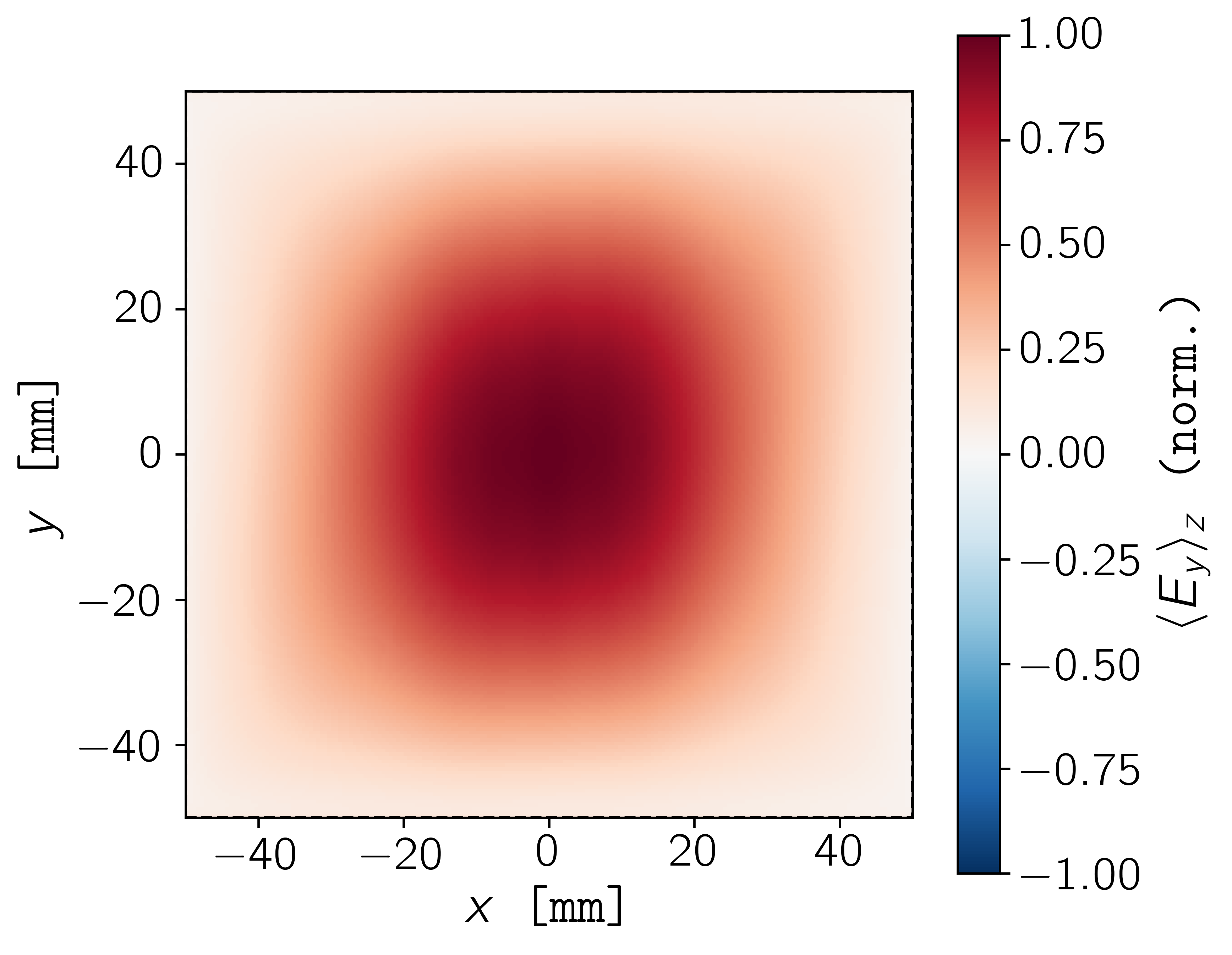}
\caption{Transverse profile of the mode on the front face of the central
plate in the $\lambda/8$ configuration, normalised to its maximum.  The
dashed contour marks the $100\times100$~mm$^2$ plate; the field extends
slightly beyond it, as the modes of a resonator of finite lateral extent
are diffractive rather than quasiplane.  The profile is single-lobed and
of one sign over the whole aperture, so that no cancellation occurs in
the numerator of Eq. \eqref{eq:eta} and the coherent projection retains
$\eta\simeq0.61$.}
\label{fig:transl8}
\end{figure}

\begin{figure}[h]
\includegraphics[width=\columnwidth]{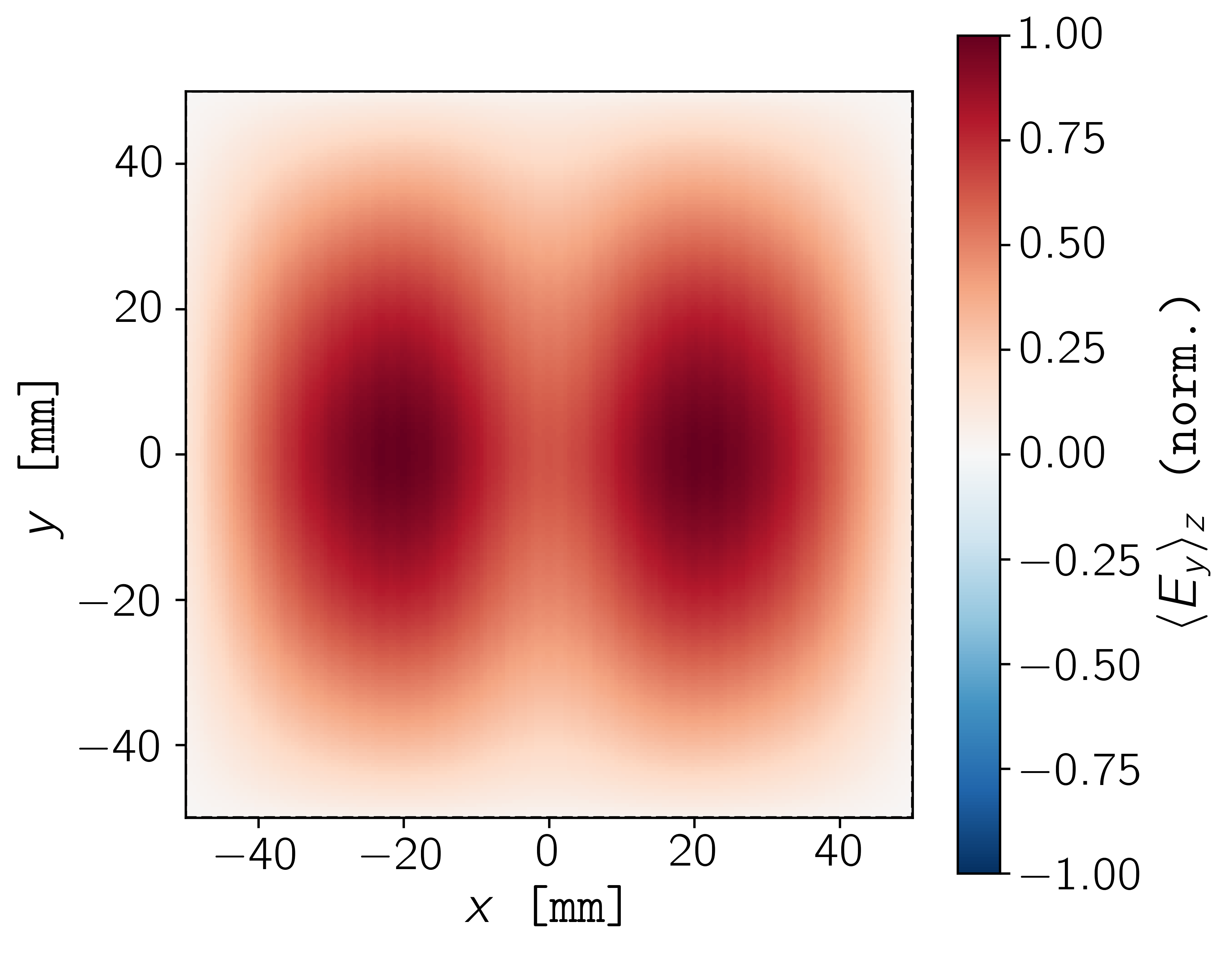}
\caption{As Fig.~\ref{fig:transl8}, for the $\lambda/2$ configuration.
Two lobes are now resolved along $\hat y$, of the same sign and separated
by a weak region of opposite sign, so that the cancellation in
\eqref{eq:eta} remains modest and $\eta\simeq0.58$.  Comparison with
Fig.~\ref{fig:transl8} shows that the transverse and longitudinal
structures vary independently: the two configurations differ by a factor
$16$ in $\CB$ and by five per cent in the mean $\eta$ of
Table~\ref{tab:eta}.}
\label{fig:transl2}
\end{figure}

Table~\ref{tab:eta} reports the transverse projection $\eta$ of
\eqref{eq:eta} on the faces of the two central plates, together with the
intensity filling factor
$A_{\rm eff}/A=\int|E|^{2}\mathrm{d}A/(\max|E|^{2}\Aphys)$, both
evaluated on the component parallel to $\bm{B}_0$.  The two are distinct
functionals of the same profile and behave differently: $\eta$ is a
coherent projection and is reduced by phase opposition between transverse
lobes, whereas $A_{\rm eff}/A$ is quadratic and cannot be.  The
distinction is the transverse counterpart of that drawn in
Sec.~\ref{sec:longitudinal} for the longitudinal direction.  In the
present geometry the two configurations differ by a factor $16$ in $\CB$
while their transverse projections agree to five per cent.

The filling factor is quoted as a diagnostic and does not enter
\eqref{eq:chain}.  Its role is to indicate whether a small $\eta$ arises
from confinement of the mode, in which case $\eta\simeq A_{\rm eff}/A$,
or from cancellation between lobes, in which case
$\eta\ll A_{\rm eff}/A$.

\begin{table}[t]
\caption{Transverse projection $\eta$ and intensity filling factor
$A_{\rm eff}/A$ on the faces of the central plates, evaluated over the
plate area.  The last row gives the mean over all planes between the port
and the mirror.}
\label{tab:eta}
\begin{ruledtabular}
\begin{tabular}{lcccc}
 & \multicolumn{2}{c}{$\lambda/8$} & \multicolumn{2}{c}{$\lambda/2$} \\
Plane & $\eta$ & $A_{\rm eff}/A$ & $\eta$ & $A_{\rm eff}/A$ \\
\colrule
P2 front & $0.631$ & $0.233$ & $0.410$ & $0.115$ \\
P2 back  & $0.615$ & $0.227$ & $0.037$ & $0.104$ \\
P3 front & $0.655$ & $0.225$ & $0.516$ & $0.123$ \\
P3 back  & $0.657$ & $0.221$ & $0.273$ & $0.103$ \\
\colrule
mean     & $0.610$ & $0.223$ & $0.583$ & $0.178$ \\
\end{tabular}
\end{ruledtabular}
\end{table}

\subsection{Dwell time at the two limits}

The independence argued in Sec.~\ref{sec:family} admits a direct test on
this hardware.  The two configurations are not two devices: they are the
same four-layer assembly, with the same $N$, $d_e$, $d_v$ and
$\varepsilon_r$, observed at two resonances.  The round-trip transit time
is therefore fixed by geometry and takes the same value in both, while
$\CB$ differs by more than an order of magnitude.

With $\tau_p=\QL/\omega_0$ and the values of Table~\ref{tab:worked}, the
photon lifetimes are $1.13$~ns and $1.31$~ns, nearly equal in absolute
time because the fourfold change in $\QL$ is very nearly cancelled by the
change in $\omega_0$.  The number of round trips executed before the
field leaves the assembly is then
$\mathcal{N}_{\mathrm{rt}}=3.30$ and $3.84$: a change of sixteen per cent,
against a sixteen\emph{fold} change in $\CB$. What this establishes is bounded, and worth stating exactly. It is not
that longitudinal coherence is irrelevant to axion excitation; it is that
longitudinal interface coherence is not the physical origin of the
resonant storage that the measurement returns. Because $\tau_{\rm rt}$ is
fixed by the geometry rather than by the mode, this comparison isolates
the quantity of interest more cleanly than a comparison across two
different stackings would.  If coherent addition contributed to resonant
storage, the change in $\CB$ would be expected to leave a corresponding
trace in $\mathcal{N}_{\mathrm{rt}}$; what is observed instead is a
residual smaller by two orders of magnitude.  Since $\QL$ is invariant
under the branch ambiguity of \eqref{eq:branches}, so are $\tau_p$ and
$\mathcal{N}_{\mathrm{rt}}$, and the comparison does not depend on that
determination.

The comparison remains two points on one device rather than a systematic
scan of $\CB$ at fixed $\mathcal{N}_{\mathrm{rt}}$, and should be read as
a bound rather than as a general proof.

\subsection{From the port-driven field to the delivered power}
\label{app:worked}

Two distinct impedances appear in this work and are kept separate here.
The $Z_0=\sqrt{\mu_0/\varepsilon_0}\simeq377\,\Omega$ of \eqref{eq:P0} is
the wave impedance of free space, which converts the axion-induced field
amplitude into a radiated Poynting flux at the mirror surface.  The
$\Zport$ of Appendix~\ref{app:oneport} is the characteristic impedance of
the guide feeding the read-out port, against which $S_{11}$ is referred.
For the oversized guide used here, operated far above cutoff, the two
coincide to better than one per cent, as noted in
Sec.~\ref{sec:transverse}.

The impedance presented to the port is not that of free space.  A bare
magnetized mirror at a distance $L$ from the port reference plane
terminates the feed in a short circuit, so that the input impedance is
reactive,
\begin{equation}
Z_{\mathrm{in}}(\omega)=j\,\Zport\tan(kL),\qquad k=\omega/c ,
\label{eq:zshort}
\end{equation}
and the reflection is essentially total.  With the stack in place the
assembly resonates, $\Gamma_0$ is real, and the input impedance reduces
to
\begin{equation}
Z(\omega_0)=\Zport\,\frac{1+\Gamma_0}{1-\Gamma_0}=\beta\,\Zport .
\label{eq:zres}
\end{equation}
Equation~\eqref{eq:zres} carries no information beyond $\beta$ itself; it
is quoted because it makes explicit that the resonant condition, and not
the free-space impedance, is what the port sees.

The power available at the mirror follows from Eq. \eqref{eq:P0} and is
independent of the stacking, since it refers to the magnetized mirror
alone,
\begin{equation}
\frac{P_0}{\Aphys}=\frac{E_0^{2}}{2Z_0}
 =2.24\times10^{-27}\,\frac{\mathrm{W}}{\mathrm{m}^{2}}\;
   C_{a\gamma\gamma}^{2}\,f_{\mathrm{DM}}
   \left(\frac{B_0}{10\,\mathrm{T}}\right)^{2}.
\label{eq:p0perarea}
\end{equation}
The remaining factors of \eqref{eq:chain} are collected in the
dimensionless combination
\begin{equation}
X\equiv\eta\,\QL\,\frac{\beta}{1+\beta} ,
\label{eq:Xdef}
\end{equation}
evaluated in Table~\ref{tab:worked}.  With $\Aphys=10^{-2}\,\mathrm{m}^{2}$
the delivered power is $\Pout=4.4\times10^{-28}\,\mathrm{W}$ at
$\lambda/8$ and $1.8\times10^{-27}\,\mathrm{W}$ at $\lambda/2$, in units
of $C_{a\gamma\gamma}^{2}f_{\mathrm{DM}}(B_0/10\,\mathrm{T})^{2}$.

\begin{table}[t]
\caption{Factor-by-factor evaluation of \eqref{eq:chain} at the two
resonances of the four-layer assembly.  $P_0$ is common to both, being a
property of the magnetized mirror rather than of the stacking.  Values of
$\beta$ are quoted in the overcoupled branch of \eqref{eq:branches}.}
\label{tab:worked}
\begin{ruledtabular}
\begin{tabular}{lcc}
 & $\lambda/8$ & $\lambda/2$ \\
\colrule
$\nu_0$ [GHz]                & $6.832$  & $24.508$ \\
$\delta/\pi$                 & $0.23$   & $0.82$   \\
$\eta$                       & $0.610$  & $0.583$  \\
$\QL$                        & $48.3$   & $201.5$  \\
$\beta$                      & $1.98$   & $2.30$   \\
$\beta/(1+\beta)$            & $0.664$  & $0.697$  \\
$\tau_p$ [ns]                & $1.13$   & $1.31$   \\
$\mathcal{N}_{\rm rt}$       & $3.30$   & $3.84$   \\
$\CB$                        & $0.050$  & $0.810$  \\
\colrule
$X$                          & $19.6$   & $81.9$   \\
$\Pout/\Aphys$ [$10^{-27}$~W\,m$^{-2}$] & $43.8$ & $183$ \\
\end{tabular}
\end{ruledtabular}
\end{table}

The ratio between the two configurations is instructive.  Writing
$\QL=\omega_0\tau_p$ and factorising,
\begin{equation}
\frac{X_{\lambda/2}}{X_{\lambda/8}}
=\underbrace{\frac{\eta_{\lambda/2}}{\eta_{\lambda/8}}}_{0.96}
\times
\underbrace{\frac{\omega_{\lambda/2}}{\omega_{\lambda/8}}}_{3.59}
\times
\underbrace{\frac{\tau_{p,\lambda/2}}{\tau_{p,\lambda/8}}}_{1.16}
\times
\underbrace{\frac{[\beta/(1+\beta)]_{\lambda/2}}
                 {[\beta/(1+\beta)]_{\lambda/8}}}_{1.05}
\approx 4.
\label{eq:Xratio}
\end{equation}
The dominant contribution is the ratio of frequencies, which enters
through $\QL=\omega_0\tau_p$: at fixed dwell time a higher resonant
frequency yields a proportionally larger loaded quality factor. That the two dwell times are so nearly equal is the numerical
counterpart of Sec.~\ref{sec:family}: at the two phase depths the slabs
present amplitude reflectivities of $0.85$ and $0.79$, so that the
assembly stores for the same length of time in both, and the whole of the
fourfold difference in $\QL$ is the frequency. The
dwell time itself, which is the physical measure of resonant storage,
changes roughly by a ten per cent, while $\CB$ changes by an order of
magnitude.  None of the four factors in \eqref{eq:Xratio} is a function
of $\CB$.

\section{Laboratory measurements}\label{app:lab}

\begin{figure}[b]
\centering
\includegraphics[width=0.5\textwidth]{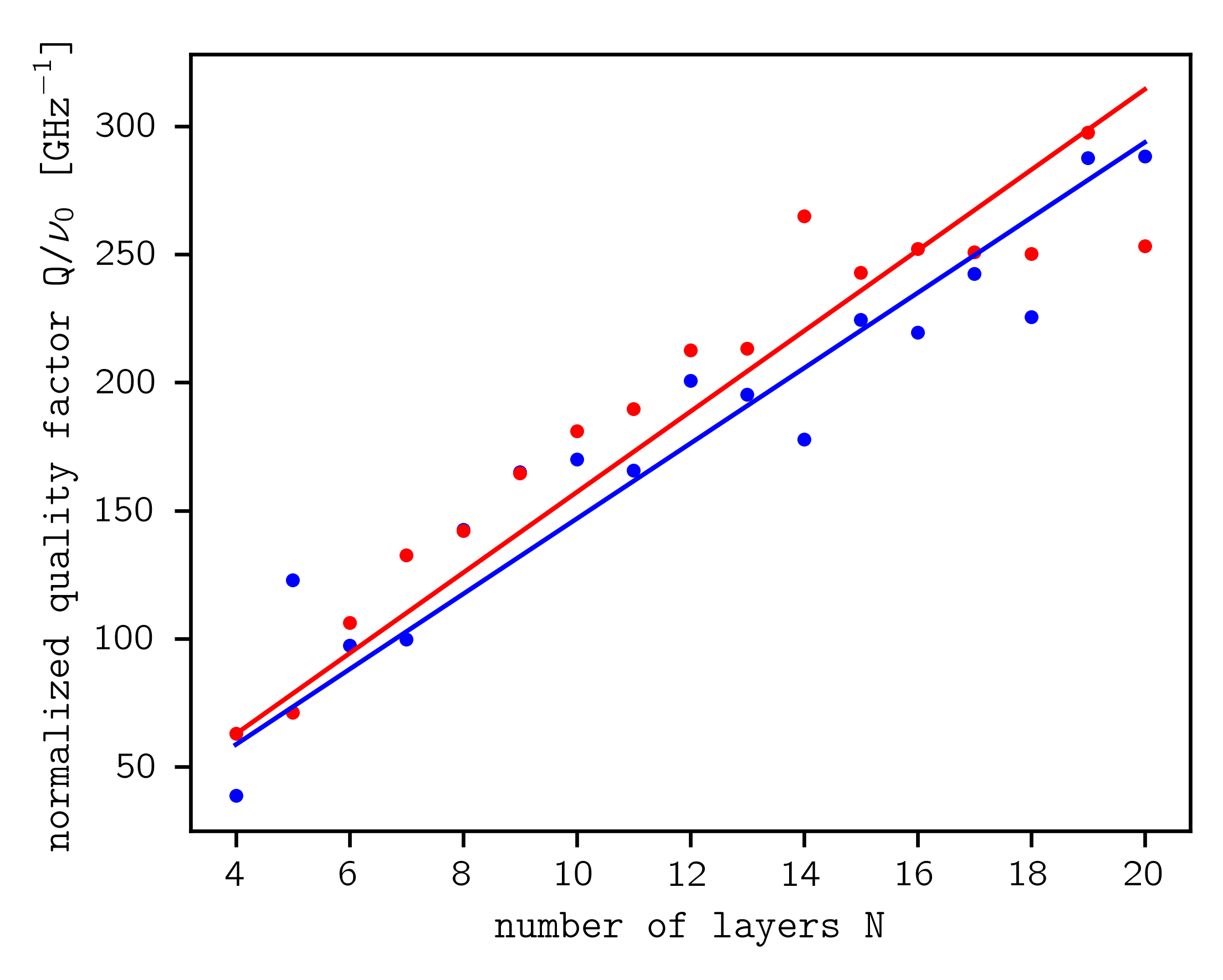}
\caption{Measured quality factors of two fixed-plate Fabry--P\'erot
haloscopes holding $N$ layers of $100\times100\times1$~mm$^{3}$
constructed with 3~mol\% yttria-stabilized zirconia, with two different
layer spacings, normalised to the resonant frequency $\nu_0$.  DUT1 is
shown in blue and DUT2 in red; the continuous lines are proportional
fits.  Both devices follow $\QL\propto N$ up to $N=20$, for even and odd
$N$ alike, with slopes that agree to within a few per cent although their
spacings, and therefore the phases with which their interfaces radiate,
differ.  Reproduced from Ref.~\cite{PhysRevD.110.072013}.}
\label{fig_4}
\end{figure}

\begin{figure}[h]
\centering
\includegraphics[width=0.5\textwidth]{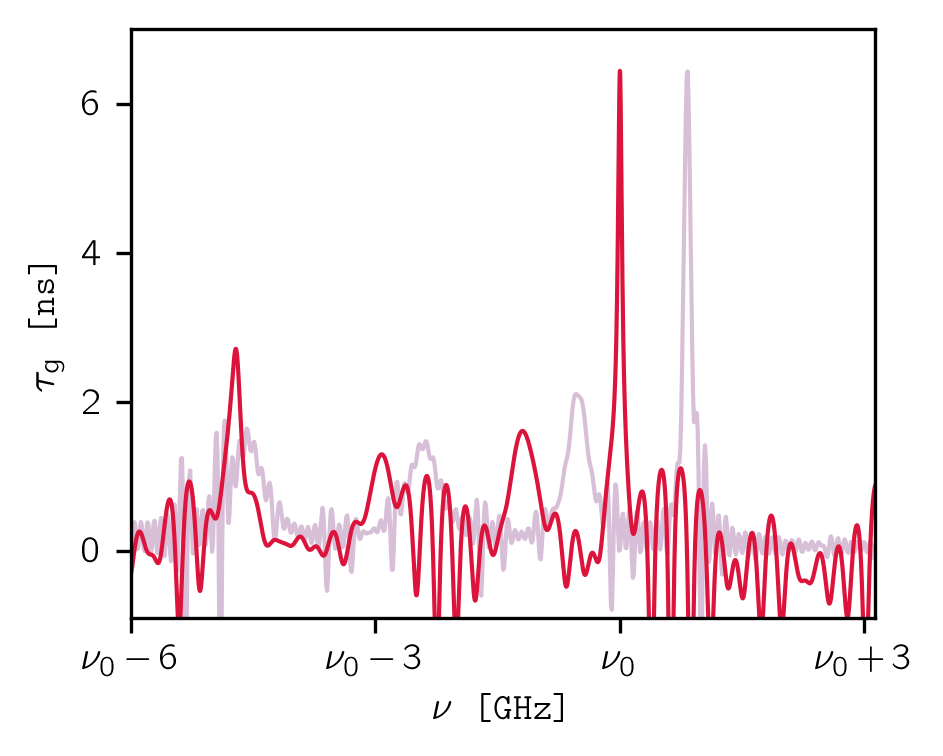}
\caption{Measured (dark red) and simulated (light red) $S_{11}$ group
delay $\tau_{\rm g}$ for the four-layer Fabry--P\'erot resonator; the
observed resonance frequency is $\nu_0$.  The group-delay maximum is
coincident with the $S_{11}$ minimum, as a resonance requires and as a
purely dissipative mismatch would not produce, and measurement and
simulation reproduce one another in the position and character of the
feature. The phase shift between simulations and measurements is compatible with small deviations ($\mathcal{O}(10\%)$) in the dielectric permittivity of the zirconia ceramic plates.
 Reproduced from Ref.~\cite{PhysRevD.110.072013}.}
\label{fig_R2}
\end{figure}

The two diagnostics of Sec.~\ref{sec:longitudinal} that require hardware
rather than simulation have been reported for the DALI
proof-of-principle apparatus~\cite{DeMiguel:2026mvi}, and the relevant
figures are reproduced here.

Figure~\ref{fig_4} realises the third diagnostic.  The linear growth of
$\QL$ with the number of layers, sustained to $N=20$ and insensitive to
the parity of $N$, identifies the feature as a resonance of the stack
rather than of the enclosure, and bounds the loss mechanism: were
dielectric loss dominant, $\QL$ would saturate at $\sim1/\tan\delta$
instead of continuing to grow.  That two devices with different spacings
share the same slope is the experimental counterpart of the argument of
Sec.~\ref{sec:family}: the quality of the resonance does not follow the
stacking.

Figure~\ref{fig_R2} realises the second.  The quantity plotted is not the
raw reflection at the port.  The response of the assembly is divided by
that of a bare mirror in the same setting, transformed to the time
domain, multiplied by a discrete prolate spheroidal window centred on the
delay of the resonator, and transformed back before the phase derivative
is taken~\cite{PhysRevD.110.072013}.  The purpose is to reject the
returns of the antenna, of the supporting structure and of the chamber,
which arrive earlier than the response of the stack and are not separable
from it in frequency; the direct term of Eq. \eqref{eq:gammasplit} is
rejected with them, so that the delay does not carry the interference
factor of Eq. \eqref{eq:taug} and is independent of the coupling factor and
of its branch.

The window is of finite width, and the filtered response is therefore not
that of an isolated pole even where the underlying resonance approaches
one: its centre and width are selected by requiring the delay to be
stationary against both, and the residual variation over that region
bounds the value.  What Fig.~\ref{fig_R2} establishes is accordingly
qualitative and insensitive to the coefficient relating $\tau_g$ to
$\QL$: a dwell time of many transits of the assembly, against
$\mathcal{N}_{\rm rt}\simeq1$ for a non-resonant reflection, with the
group-delay maximum coincident in frequency with the $S_{11}$ minimum.

\section{Relation to the transfer-matrix formalism}\label{app:tm}

Multilayer stacks are conventionally described with transfer matrices,
which return a single boost amplitude in which resonant storage and
coherent addition are not separated~\cite{Millar:2016cjp}.  The present
work factorises the delivered power into $\mathcal{F}_m$ and $\QL$.  The
two descriptions are not alternatives: the second is the
isolated-resonance limit of the first.  This appendix derives that
correspondence, states the condition under which it holds, and identifies
what the factorisation does and does not separate.

\subsection{The round-trip operator}

Consider an assembly of $N$ layers terminated by a mirror.  Let
$\mathbf{a}$ collect the complex amplitudes of the waves travelling in
either direction at each of the $2N$ dielectric faces and at the mirror,
and let $\mathsf{S}$ denote the operator that advances this state by one
complete transit: the waves propagate across the gaps and the layers,
reflect and transmit at each face, and return to their starting points,
\begin{equation}
\mathbf{a}\;\longmapsto\;\mathsf{S}\,\mathbf{a} .
\label{eq:roundop}
\end{equation}
Every property of the geometry enters $\mathsf{S}$ and nothing else does:
the Fresnel coefficients of each face, fixed by $\varepsilon_r$, and the
phases accumulated in traversing each gap and each layer, fixed by the
thicknesses and the frequency.  The steady-state field is the sum of the
contributions of all transits,
\begin{equation}
\mathbf{a}_{\rm tot}
=\left(\mathbb{1}+\mathsf{S}+\mathsf{S}^{2}+\cdots\right)\mathbf{a}_0
=\left(\mathbb{1}-\mathsf{S}\right)^{-1}\mathbf{a}_0 ,
\label{eq:resolvent}
\end{equation}
so that the resolvent $(\mathbb{1}-\mathsf{S})^{-1}$ is the accumulated
effect of multiple reflection, resummed to all orders.

\subsection{Why the resummation is scalar only in the simplest case}

For a resonator formed by two terminations enclosing a single optical
path, $\mathsf{S}$ reduces to a number and the resummation is the
familiar scalar
\begin{equation}
\left(\mathbb{1}-\mathsf{S}\right)^{-1}\;\longrightarrow\;
\frac{1}{1-r_1r_2\,e^{2i\delta}} ,
\label{eq:airy}
\end{equation}
with $r_{1,2}$ the reflectivities of the terminations.
Equation~\eqref{eq:airy} is written to display the structure; for a
dielectric layer of finite thickness before a mirror the exact closed
form is more involved, since the internal structure of the layer cannot
be replaced by a single effective reflector, and is given in
Ref.~\cite{Millar:2016cjp}.  The essential feature survives that
refinement: the resonance resides in a denominator that becomes small
when the round trip returns in phase.

For $N>1$ this scalar form is not available.  Closed paths exist between
every pair of faces and between each face and the mirror, and multiple
scattering couples them, so that the enhancement experienced by the
emission of one interface differs from that experienced by another and no
common factor can be extracted from the sum over interfaces.  The
resummation Eq. \eqref{eq:resolvent} remains exact, but it is a matrix
inverse rather than a scalar.  This is the formal reason why the
transfer-matrix description delivers a single number and offers no
natural decomposition into storage and coherent addition.

\subsection{Spectral decomposition and the emergence of a mode}

A decomposition does exist, but it requires diagonalising $\mathsf{S}$
rather than factorising it.  Let
\begin{equation}
\mathsf{S}\,\mathbf{v}_n=\lambda_n\,\mathbf{v}_n
\label{eq:eigen}
\end{equation}
define its eigenvectors and eigenvalues.  Each $\mathbf{v}_n$ is a
distribution of amplitude over the faces which reproduces itself after
one transit, changed only by the complex factor $\lambda_n$; the modulus
$|\lambda_n|$ states what fraction survives, and $\arg\lambda_n$ the
phase with which it returns. An open and lossy assembly has no reason to
make $\mathsf{S}$ normal, so its right and left eigenvectors differ,
$\mathsf{S}\mathbf{v}^{R}_n=\lambda_n\mathbf{v}^{R}_n$ and
$\mathbf{v}^{L\dagger}_n\mathsf{S}=\lambda_n\mathbf{v}^{L\dagger}_n$,
normalised biorthogonally as
$\mathbf{v}^{L\dagger}_m\mathbf{v}^{R}_n=\delta_{mn}$.  In that basis the
resolvent is diagonal,
\begin{equation}
\left(\mathbb{1}-\mathsf{S}\right)^{-1}
=\sum_n\frac{\mathbf{v}^{R}_n\mathbf{v}^{L\dagger}_n}{1-\lambda_n} ,
\label{eq:spectral}
\end{equation}
each self-reproducing distribution acquiring its own scalar enhancement.

A resonance is the statement that one eigenvalue, and only one,
approaches unity: the corresponding distribution returns in phase and
survives the transit almost intact.  Writing
$\lambda_{\rm res}=1-\epsilon$ with $|\epsilon|\ll1$, that term of
\eqref{eq:spectral} is of order $1/\epsilon$ while the others remain of
order unity, and the sum collapses to
\begin{equation}
\left(\mathbb{1}-\mathsf{S}\right)^{-1}
\;\longrightarrow\;
\frac{\mathbf{v}_{\rm res}\mathbf{v}_{\rm res}^{\dagger}}{\epsilon} .
\label{eq:collapse}
\end{equation}
The two surviving objects are those used throughout this work.  The
eigenvector $\mathbf{v}_{\rm res}$ is the distribution that reproduces
itself, which is the standing wave $e_m(z)$ of
Sec.~\ref{sec:sourceindep}, and the projection of the source onto it is
$\mathcal{F}_m$ of \eqref{eq:Fsum}. The field profile is the right eigenvector; that the source couples to
that same profile is the content of the reciprocity argument of
Sec.~\ref{sec:sourceindep}, and does not rest on the two coinciding. The scalar $1/\epsilon$ is the reciprocal of the fractional loss per
transit, and is the resonant enhancement.  Since $\mathsf{S}$ advances
complex amplitudes, $\epsilon$ is the fractional loss of
\emph{amplitude}, and the stored energy, going as the square, decays
twice as fast.  The number of round trips the energy survives is
therefore
\begin{equation}
\mathcal{N}_{\rm rt}=\frac{1}{2\epsilon}=\frac{\tau_p}{\tau_{\rm rt}},
\qquad
\tau_p=\frac{\QL}{\omega_0} ,
\label{eq:epsQ}
\end{equation}
in the convention of Eq. \eqref{eq:renkQ}, where the survival factor $V$ is
that of the energy.  The factor two is the same one that separates Eq.
\eqref{eq:taug} from Eq.\eqref{eq:taugated} in Appendix~\ref{app:oneport}.
The boost amplitude therefore reduces, near an isolated resonance, to
\begin{equation}
B\;\longrightarrow\;
\underbrace{\left[\sum_i s_i\,e_m(z_i)\right]}_{\mathcal{F}_m}
\;\times\;
\underbrace{\frac{1}{\epsilon}}_{\propto\;\QL} ,
\label{eq:Bfactor}
\end{equation}
which is the factorisation adopted in Sec.~\ref{sec:sourceindep}. It is not an additional assumption but
the resonant limit of the same calculation. The identification
$\mathcal{N}_{\rm rt}=1/2\epsilon$ is the leading term of
$\mathcal{N}_{\rm rt}=-1/\ln|\lambda_{\rm res}|^{2}$ for $|\epsilon|\ll1$;
away from the exact resonance $\epsilon$ acquires a phase and the reading as a number of transits is approximate.

\subsection{Consistency with the single-layer result}

The construction can be checked against the one case for which the
transfer-matrix result is available in closed form.  For a mirror with a
single dielectric layer of refractive index $n$,
Ref.~\cite{Millar:2016cjp} observes that near resonance the boost
amplitude takes the form of a driven damped oscillator,
$B\propto(\omega-\omega_R+i\Gamma/2)^{-1}$, and obtains the quality
factor by precisely the argument used above. The fraction of the stored
field that escapes per transit is the transmittance of the layer,
$|T|^{2}\sim n^{-2}$, so that the number of transits, and hence the
enhancement, is $|T|^{-2}\sim n^{2}$.  The explicit result is
$\Gamma=2\omega_R/\pi n^{2}$ and $Q=\pi n^{2}/2$.  The identification Eq.
\eqref{eq:epsQ} is therefore the same statement, generalised from one
layer to $N$.

\subsection{What the factorisation does and does not separate}

Two cautions follow. The first concerns the relation between the two factors.  That
$\mathcal{F}_m$ and $\QL$ occupy different places in Eq. \eqref{eq:Bfactor}
does not make them numerically unconstrained: both are functions of the
same geometry.  In the quarter-wave configuration of the single-layer
cavity the boost is produced almost entirely by storage, and the peak
boost factor and the quality factor are proportional; in the half-wave
configuration the layer is transparent, its internal reflections
cancelling, no field accumulates, and
the boost amplitude reduces to the coherent sum of the surface emissions
with no resonant factor at all.  What Eq. \eqref{eq:Bfactor} establishes is
that the two enter through distinct functionals---one linear in the field
and weighted against the source, the other a ratio of quadratic
functionals---so that $\QL$ cannot register the relative phases of the
interface contributions.  Which mechanism dominates is a property of the
configuration, and Sec.~\ref{sec:family} shows that the two are
controlled by different functions of the phase depth.

The second concerns the partition itself.  A multiple-scattering
expansion of Eq. \eqref{eq:resolvent} separates the response into a term in
which each interface radiates and propagates freely to the exit, and a
remainder containing every path with at least one internal reflection.
It is tempting to identify the first with coherent addition and the
second with resonance.  That is a different partition from the one in Eq.
\eqref{eq:Bfactor}. There, $\mathcal{F}_m$ is evaluated against $e_m$,
whose shape already embodies the interference of the multiply reflected
waves, since the positions of its nodes would not exist without them.
The two decompositions cut the same quantity along different surfaces and
their terms are not in correspondence.  In particular
$\sum_i s_i e_m(z_i)$ is not the single-pass sum of the surface emissions
and should not be read as such.

Finally, Eq. \eqref{eq:collapse} requires one eigenvalue of $\mathsf{S}$ to
dominate the others.  This is a sharper form of the isolated-resonance
requirement invoked in Sec.~\ref{sec:sourceindep}. What must be separated
is not merely the resonant frequencies but the eigenvalues of the
round-trip operator.  Where two approach unity together, no single
standing wave describes the field and the factorisation ceases to apply,
while the transfer-matrix description remains valid. That is the price
of the factorisation, and the reason both formalisms are useful.

\bibliography{apssamp.bib}

\end{document}